\documentclass[reprint, amsmath,amssymb,aps]{revtex4}
\usepackage{hyperref}
\usepackage{graphicx}
\usepackage{epstopdf}
\usepackage{amsmath,amssymb,amsfonts,amsthm,latexsym,stmaryrd,mathtools}
\usepackage{dsfont}
\usepackage{esint}
\usepackage[usenames, dvipsnames]{color}
\usepackage{subfig}
\def\nn{\nonumber}
\def\be{\begin{equation}}
\def\ee{\end{equation}}
\def\ba{\begin{eqnarray}}
\def\ea{\end{eqnarray}}
\allowdisplaybreaks[4]

\begin{document}

\title{Nonclassical cosmological dynamics in the low-energy limit of loop quantum scalar-tensor theory}

\author{Yu Han}
 \email{hanyu@xynu.edu.cn}
 \affiliation{College of Physics and Electrical Engineering, Xinyang Normal University, 464000 Xinyang, China}

\begin{abstract}	
In previous work, we showed that in loop quantum cosmology of scalar-tensor theory (STT) with the holonomy correction the background equations of motion in the Jordan frame have two branches, i.e., the $b_{+}$ branch and the $b_{-}$ branch. In the low-energy limit,  the $b_{+}$ branch of the equations of motion reproduce the equations of motion of classical STT while the $b_{-}$ branch of equations of motion do not reproduce the classical equations. In this paper, we investigate cosmological dynamics in an expanding universe whose background is described by the the $b_{-}$ branch of equations of motion of STT, and we especially focus on the dynamics of the perturbations in the low-energy limit because it is most relevant to the current observational range.
First, we show that the low-energy limit of the $b_{-}$ branch of equations of motion can be a stable attractor in the expansion phase of a universe. Then, we find a low-energy effective Hamiltonian on the spatially flat Friedmann-Robertson-Walker background. The background part of this Hamiltonian can yield the low-energy limit of the $b_{-}$ branch of equations, and this Hamiltonian consists of constraints whose constraint algebra is different from the classical case but also closed up to arbitrary order of perturbations.  Remarkably, we find that this Hamiltonian can be transformed into the Hamiltonian of the Einstein frame by field redefinitions different from the classical case. Moreover, we also develop the linear cosmological perturbation theory and apply it to study the slow-roll inflation in this context. Finally, we study a specific model of STT. In this model, a contracting universe described by classical STT in the remote past can pass through the bounce and evolve into an expanding universe whose background dynamics is described by the $b_{-}$ branch of equations of motion. It is also shown that the slow-roll inflation can take place in this case, and the spectral indices of the slow-roll inflation agree well with the observations. The results in this paper indicate that there exists an alternative consistent theory which is different from the classical theory in the low-energy limit of loop quantum STT.
\end{abstract}

\maketitle


\section{Introduction}
In the past few decades, scalar-tensor theory has been seriously considered as a natural generalization of general relativity by many researchers in cosmology. Recently, astrophysical observations indicate that the predictions about the slow-roll inflation in some specific models of STT agree very well with the  observational data \cite{Ade:2014,Akrami:2018}, which triggers renewed research interest of various inflationary models in STT (see, for instance, Refs. \cite{Kallosh:2014,Giudice:2014,Kaneta:2018,Ferreira:2018}).  Nevertheless, the quantum gravity effects in STT which may also leave footprints during the slow-roll inflation have been neglected to a large extent yet.  Fortunately, the recent development of loop quantum cosmology (LQC) allows us to preliminarily investigate the quantum gravity effects in STT.

LQC is a tentative quantum cosmology theory which implements the quantization techniques of loop quantum gravity on the reduced phase space in the cosmological case. Among the several characteristic quantum corrections of LQC, the holonomy correction in which the holonomy of the connection around a given square is treated as the fundamental variable is extensively studied in the literature. The most important effect caused by the holonomy correction is that the cosmological singularity is replaced by a bounce. This result holds irrespective  of the choice of quantization prescription; i.e., the bounce exists no matter we choose the standard holonomy quantization prescription in which the Euclidean term and Lorentz term of the Hamiltonian constraint are treated on the same footing \cite{Ashtekar:2006,Ashtekar:2011} or the modified holonomy quantization in which the Euclidean term and Lorentz term are treated differently in the way which mimics the quantization prescription of the Hamiltonian constraint of loop quantum gravity \cite{Yang:2009,Dapor:2018,Assanioussi:2018}. Despite the uniform existence of bounce, a different choice of holonomy quantization prescription can lead to essentially different behaviors in the effective dynamics. For instance, in the effective dynamics of the minimally coupled models using the standard holonomy quantization, a collapsing classical universe in the remote past is connected with an expanding classical universe in the asymptotic future via the bounce, while in the case with the modified holonomy quantization, a contracting de-Sitter universe in the remote past is connected with an expanding classical universe in the asymptotic future via the bounce.

In LQC of STT, the holonomy correction can been studied in the Einstein frame or in the Jordan frame. In the Einstein frame, the holonomy of the conformally transformed connection is quantized, while in the Jordan frame the holonomy of the connection itself is directly quantized.
 For the sake of simplicity, in the literature, the standard or the modified holonomy quantization has been frequently studied in the Einstein frame in some specific models of STT \cite{Artymowski:2012,Amoros:2014,Odintsov:2014,Jin:2018,Haro:2018,Li:2018a,Li:2018b}. In the Jordan frame, the standard holonomy quantization was first applied to Brans-Dicke theory in Refs. \cite{Zhang:2013,Artymowski:2013,Chen:2018} and  extended to STT in Ref. \cite{Han:2019} on the spatially flat Friedmann-Robertson-Walker (FRW) background.  In the Einstein frame, the effective cosmological dynamics mimics that of the minimally coupled case, while in the Jordan frame the effective dynamics turns out to be much different from that in the Einstein frame. A key difference pointed out in Ref. \cite{Han:2019} is that the background equations of motion of STT in the Jordan frame have two branches, namely the $b_{+}$ branch and the $b_{-}$ branch. In the low-energy limit, the $b_{+}$ branch of equations of motion can reproduce the equations of motion of classical STT while the $b_{-}$ branch of equations of motion cannot reproduce the classical equations. The two branches of equations of motion can be connected to each other by the quantum bounce, which means if a contracting universe described by the $b_{+}$ branch of equations of motion passes through the quantum bounce, it will evolve into an expanding universe whose background dynamics is described by the $b_{-}$ branch of equations of motion and vice versa.

  Assuming that the background dynamics of the current expanding Universe is described by the $b_{-}$ branch of equations of motion of STT, it is natural to ask whether we can obtain more information relevant to observations in this context. To get a clear and definite answer to this question, we have to investigate both the background and perturbation dynamics. Since the present observations of slow-roll inflation have accumulated comparatively rich and accurate data, provided that the slow-roll inflation can take place in this case, we can investigate perturbation effects during the slow-roll inflation to obtain useful observational information. In order to do this, we have to use the cosmological perturbation theory. Considering that the range of the comoving wave numbers that the current observations can cover is $k\geq 0.002$ Mpc$^{-1}$ and for any wave number in this range the energy density at an instant of the horizon crossing during the slow-roll inflation is generally lower than the critical energy density of LQC by many orders of magnitude, therefore, we may only focus on the perturbation effects of these wave numbers in the low-energy limit for the sake of simplicity. However, since the spacetime background described by the low-energy limit of the $b_{-}$ branch of equations is nonclassical, the evolution of cosmological perturbations propagating on the background is nonclassical too.  Hence, in this case, the classical perturbation theory is no longer applicable, and we have to construct an alternative cosmological perturbation theory.
  It is only after a consistent perturbation theory is established that we can possibly draw reliable information from perturbations during the slow-roll inflation. To summarize, in this paper, we aim to construct a consistent STT on the spatially flat FRW background in the low-energy limit which is different from the classical theory. This theory and the classical theory can be viewed as two different limiting cases of LQC of STT.

The structure of this paper is as follows. In Sec. II, we review the two branches of background equations of motion of STT in the Jordan frame and analyze the dynamical properties in the low-energy limit of STT. In Sec. III, we obtain the background Hamiltonian which can yield the low-energy limit of the $b_{-}$ branch of equations of motion. Using the background Hamiltonian and the approach of anomaly-free algebra, we obtain the effective Hamiltonian in the low-energy limit on the spatially flat FRW background.  We also introduce the field redefinitions that can transform the Hamiltonian of the Jordan frame into  the Hamiltonian of the Einstein frame. In Sec. IV, we expand the Hamiltonian to the second order of perturbations and derive the cosmological perturbation equations of the gauge invariant perturbed variables; we also discuss the issue of causality with regard to the perturbation equations. In Sec. V, we solve the perturbation equation under the slow-roll approximation and obtain the spectral indices in the Jordan frame; the results are compared with those in the Einstein frame. In Sec. VI, using the results obtained in previous sections, we study the cosmological dynamics of a specific model of STT. In the last section, we conclude and make some remarks.

\section{Background dynamics of LQC of STT}
In this section, we first review the two branches of equations of motion of STT in the cosmological case, and then, we focus on the issue of stability of dynamics in the low-energy limit of STT.
\subsection{Background equations of motion of STT}
In this subsection, we briefly review some results obtained in Ref. \cite{Han:2019}.

The classical action of STT we use is given by
 \ba
S=&&\int_{\mathcal{M}}d^4x\sqrt{|\det(g)|}\bigg[\frac{1}{2\kappa}F(\phi)R^{(4)}
-\frac{1}{2}K(\phi)(\partial^{\mu}\phi)\partial_{\mu}\phi-V(\phi)\bigg],
\label{STTaction}
\ea
  in which $\mathcal{M}$ is the four-dimensional spacetime manifold, $\kappa=8\pi G$, and  $F(\phi)$, $K(\phi)$ are dimensionless coupling functions of the scalar field, and $V(\phi)$ is the potential.

On the spatially flat FRW background,  the background equations of motion of STT with the holonomy correction in the Jordan frame are as follows;
\ba
&&\bigg(F\big(\bar{\phi}\big)H+\frac{1}{2}\dot{F}\big(\bar{\phi}\big)\cos b\bigg)^2
=\frac{\kappa}{3}\rho_e\bigg(1-\frac{\rho_e}{\rho_c}\bigg),\label{qFr}\\
&&\ddot{\bar{\phi}}+3H\dot{\bar{\phi}}
+\frac{1}{2}\frac{\dot{G}\big(\bar{\phi}\big)}{G\big(\bar{\phi}\big)}\dot{\bar{\phi}}-\frac{(3\cos b-1)F'\big(\bar{\phi}\big)V\big(\bar{\phi}\big)
-F\big(\bar{\phi}\big)V'\big(\bar{\phi}\big)}{G\big(\bar{\phi}\big)}=0,\label{qKG}\\
&&\cos^2b=1-\frac{\rho_e}{\rho_c}, \label{cos2b}
\ea
where $\bar{\phi}$ is the background component of $\phi$.  $H\equiv\frac{\dot{a}}{a}$ is the Hubble parameter in which $a$ denotes the scale factor. Throughout this paper, we use an overdot to denote the derivative with respect to the proper time $t$ and the prime to denote the derivative with respect to $\phi$;  i.e., $\dot{\bar{\phi}}\equiv\frac{d\bar{\phi}}{d t}$, $V'\big(\bar{\phi}\big)\equiv\frac{dV(\bar{\phi})}{d \bar{\phi}}$. In Eqs. (\ref{qFr}) and (\ref{qKG}), $\cos b$ is a component of the holonomy function,  and $\rho_e$ denotes the effective energy density of the scalar field defined by
\ba
\rho_e\equiv\frac{G\big(\bar{\phi}\big)}{2}\big(\dot{\bar{\phi}}\big)^2
+F\big(\bar{\phi}\big)V\big(\bar{\phi}\big),\quad G\big(\bar{\phi}\big)\equiv\frac{3}{2\kappa}\Big(F'\big(\bar{\phi}\big)\Big)^2
+F\big(\bar{\phi}\big)K\big(\bar{\phi}\big),\label{rhoe}
\ea
and $\rho_c\equiv \frac{3}{\Delta\kappa\gamma^2}$ is the critical energy density in LQC which depends on the Barbero-Immirzi parameter $\gamma$ and the smallest quantum, $\Delta$, of the area in loop quantum gravity.

 From the quantum effective Friedmann equation (\ref{qFr}), the Klein-Gordon equation (\ref{qKG}), and the constraint (\ref{cos2b}), we can derive the equations of motion of the other background variables. For instance, the evolution of $\cos b$ satisfies
\ba
\dot{\cos b}=\epsilon \frac{\sqrt{3\kappa\rho_e}}{2\rho_c}\frac{G\big(\bar{\phi}\big)}{F\big(\bar{\phi}\big)}
\Big(\dot{\bar{\phi}}\Big)^2
-\frac{3}{2}\frac{\rho_e}{\rho_c}\frac{\dot{F}\big(\bar{\phi}\big)}{F\big(\bar{\phi}\big)},\label{dotcosb}
\ea
where $\epsilon\equiv \text{sgn}\Big(HF(\bar{\phi})\cos b+\frac{1}{2}\dot{F}\big(\bar{\phi}\big)\cos^2 b\Big)$, and the evolution of the Hubble parameter satisfies
\ba
&&F\big(\bar{\phi}\big)\dot{H}\cos b-\dot{F}\big(\bar{\phi}\big)H\bigg(\frac{3}{2}\cos 2b-\cos b\bigg)+\frac{1}{2}\Big[\dot{F}\big(\bar{\phi}\big)\dot{\cos b}+\kappa K\big(\bar{\phi}\big)\Big(\dot{\bar{\phi}}\Big)^2\cos 2b+\ddot{F}\big(\bar{\phi}\big)\cos b\Big]\cos b=0.\label{qRaeq}
\ea

From Eq. (\ref{cos2b}), we have
\ba
(\cos b)_{\pm}=\pm\sqrt{1-\frac{\rho_e}{\rho_c}}.\label{cosb}
\ea
We call the equations of motion with positive $\cos b$ the $b_{+}$ branch of equations of motion and the equations with negative $\cos b$ the $b_{-}$ branch of equations of motion. In the low-energy limit $\frac{\rho_e}{\rho_c}\rightarrow0$, the equations of motion of classical STT can be reproduced by
the $b_{+}$ branch of equations of motion but not by the $b_{-}$ branch of equations of motion. From Eq. (\ref{qFr}), we learn that $H=0$ when $\cos b=0$, which means a universe bounces or recollapses at $\cos b=0$ if $\dot{H}\neq0$. Note that in classical STT the bounce or recollapse of a universe can also take place under very special conditions \cite{Boisseau:2015}. To distinguish from the classical bounce or recollapse, in the following, we call the bounce or recollapse associated with $\cos b=0$ the quantum bounce or recollapse. The two branches of equations of motion are connected with each other if a universe undergoes the quantum bounce or recollapse during the evolution, and whether this condition can be satisfied should be checked case by case.

In LQC of STT,  the background evolution of a contracting or expanding universe can be described by either branch of equations of motion. In particular, supposing that in a specific model of STT the quantum bounce can take place during the evolution, then it is possible for a contracting universe described by the $b_{+}$ branch of equations of motion  to pass through the bounce and evolve into an expanding universe described by the $b_{-}$ branch of equations of motion.

\subsection{Stability analysis in the low-energy limit}
 Assuming that the background of an expanding universe is described by the $b_{-}$ branch of equations of motion of STT, to ensure that a universe can evolve to a low-energy state, the low-energy limit of the $b_{-}$ branch should be an attractor in the phase space.  In this subsection, we show that such attractor can exist under certain conditions.

 From Eq. (\ref{cos2b}), we have
 \ba
 \sin^2 b=\frac{\rho_e}{\rho_c},\label{sin2b}
 \ea
and substituting it into Eq. (\ref{qFr}), we obtain
\ba
H=\frac{1}{F\big(\bar{\phi}\big)}\bigg(\sqrt{\frac{\kappa\rho_c}{3}}\sin b-\frac{1}{2}\dot{F}\big(\bar{\phi}\big)\bigg)\cos b.\label{eqH}
\ea

Defining $\chi\equiv\dot{\bar{\phi}}$ and using Eq. (\ref{qKG}), we obtain the following two-dimensional dynamical system in the phase space,
\ba
\frac{d\bar{\phi}}{dt}&=&\chi,\nn\\
\frac{d\chi}{dt}&=&-3H\chi-\frac{1}{2}\frac{G'\big(\bar{\phi}\big)}{G\big(\bar{\phi}\big)}\chi^2+\frac{(3\cos b-1)F'\big(\bar{\phi}\big)V\big(\bar{\phi}\big)-F\big(\bar{\phi}\big)V'\big(\bar{\phi}\big)}{G\big(\bar{\phi}\big)},\label{dpsidt}
\ea
in which  $\cos b$ and $H$ are understood as functions of $\bar{\phi}$ and $\dot{\bar{\phi}}$ (up to the signs of $\sin b$ and $\cos b$) through Eqs. (\ref{cos2b}) and  (\ref{eqH}).

If the potential $V\big(\bar{\phi}\big)$ has a minimum at $\bar{\phi}=\bar{\phi}_o$ and $V(\bar{\phi}_o)=0$,
the dynamical system will have a fixed point $\big(\bar{\phi}=\bar{\phi}_o,\chi=0\big)$ around which $\frac{\rho_e}{\rho_c}\rightarrow0$. Moreover, if
$F\big(\bar{\phi}_o\big)>0$ and $K\big(\bar{\phi}_o\big)>0$,
the fixed point $(\bar{\phi}_o,0)$ will be either a sink of the dynamical system or a source of the dynamical system, to which we give a detailed explanation as follows:
Using Eqs. (\ref{rhoe}) and (\ref{sin2b}), we find that $\sin b$ is either positive definite or negative definite around the fixed point $\big(\bar{\phi}_o,0\big)$; i.e., the sign of $\sin b$ never changes during the evolution near this point. Now, we analyze four different kinds of behavior in the vicinity of this fixed point.

(i) If $\sin b\rightarrow0_{+}$ and $\cos b\rightarrow1$, from Eq. (\ref{eqH}) we find that $H>0$ near the fixed point, which corresponds to an expanding universe described by classical STT. In this case, we can show that there exists a Lyapunov function $f\equiv\frac{\rho_e}{F^3(\bar{\phi})}$ which is positive definite and decreases monotonically with respect to the proper time near this fixed point,
\ba
\frac{df}{dt}&=&-\frac{3\rho_c\sin b}{F^4\big(\bar{\phi}\big)}\bigg[\sqrt{\frac{\kappa}{3\rho_c}}G\big(\bar{\phi}\big)
\big(\dot{\bar{\phi}}\big)^2\cos b+\dot{F}\big(\bar{\phi}\big)(1-\cos b)\sin b\bigg]\nn\\
&\simeq&-\sqrt{3\kappa\rho_c}\frac{G\big(\bar{\phi}\big)}{F^4\big(\bar{\phi}\big)}
\big(\dot{\bar{\phi}}\big)^2\sin b<0;\label{dfdt}
\ea
thus, this fixed point is asymptotically stable and an attractor, or a sink, of the dynamical system.

(ii) If $\sin b\rightarrow0_{-}$ and $\cos b\rightarrow1$, we have $H<0$ near the fixed point, which corresponds to a contracting universe described by classical STT. In this case,  this fixed point is asymptotically unstable and a source of the system.

(iii) If $\sin b\rightarrow0_{-}$ and $\cos b\rightarrow-1$, we have $H>0$ near the fixed point, which corresponds to an expanding universe described by the low-energy limit of the $b_{-}$ branch of equations of motion, and we can also find a Lyapunov function $\tilde{f}\equiv F^3(\bar{\phi})\rho_e$ which is positive definite and decreases monotonically with respect to the proper time around this fixed point,
\ba
\frac{d\tilde{f}}{dt}&=&-3\rho_cF^2\big(\bar{\phi}\big)\sin b\bigg[\sqrt{\frac{\kappa}{3\rho_c}}G\big(\bar{\phi}\big)\big(\dot{\bar{\phi}}\big)^2\cos b-\dot{F}\big(\bar{\phi}\big)(1+\cos b)\sin b\bigg]\nn\\
&\simeq&\sqrt{\frac{3\kappa}{\rho_c}}F^2\big(\bar{\phi}\big)
G\big(\bar{\phi}\big)\big(\dot{\bar{\phi}}\big)^2\sin b<0.\label{dfdt}
\ea
In this case, the fixed point $\big(\bar{\phi}_o,0\big)$ is also asymptotically stable and a sink of the system.

(iv) If $\sin b\rightarrow0_{+}$ and $\cos b\rightarrow1$, we have $H<0$ near the fixed point, which corresponds to a contracting Universe described by the low-energy limit of the $b_{-}$ branch of equations of motion. In this case, the fixed point $\big(\bar{\phi}_o,0\big)$ is asymptotically unstable and a source of the system.

Hence, we conclude that the fixed point $\big(\bar{\phi}_o,0\big)$ can also be a local attractor of the dynamical system in the low-energy limit with $\cos b\rightarrow-1$ . Note that there might be other local attractors corresponding to different fixed points in the low-energy limit with $\cos b\rightarrow-1$  in a specific model of STT.

In addition, if $\bar{\phi}_o$ is a global minimum of $V\big(\bar{\phi}\big)$ at which $V\big(\bar{\phi}_o\big)=0$ and the coupling functions satisfy $F\big(\bar{\phi}\big)>0$, $K\big(\bar{\phi}\big)>0$ for an arbitrary value of $\bar{\phi}$, then from the definition of $\rho_e$ we know that $\rho_e$ can only vanish at $(\bar{\phi}_o,0)$. Since $(\bar{\phi}_0,0)$ is a fixed point of the phase space, the sign of $\sin b$ never changes during the evolution. Thus, the phase space can be divided into two disconnected sectors by the sign of $\sin b$ . From Eq. (\ref{eqH}), we find that in the sector $\sin b<0$ a contracting Universe is described by the $b_{+}$ branch , and an expanding Universe is described by the $b_-$ branch of equations. Since the fixed point $(\bar{\phi}_o,0)$ with $\cos b=1$ is a source of the system, we can set the initial condition in the asymptotic past where a contracting universe is described by the classical STT. Moreover, if there are no limit circles and no other sinks exists except for the attractor $(\bar{\phi}_o,0)$ with $\cos b=-1$ in the $\sin b<0$ sector, it is possible for the solutions of equations of motion starting from the low-energy limit with $\cos b\rightarrow1$ to stably approach the low-energy limit with $\cos b\rightarrow-1$. In other words, in the $\sin b<0$  sector it is possible for a contracting universe described by classical STT in the asymptotic past to pass through the quantum bounce and approach the low-energy limit with $\cos b\rightarrow-1$ in the expansion phase of a universe in the asymptotic future. In Sec. VI, we give a concrete description of such evolution in a specific model of STT.

In the following sections of this paper, we aim to explore the background and perturbation dynamics of an expanding universe whose background is described by the $b_-$ branch of equations. As explained in Sec. I, we mainly focus on the perturbation dynamics in the low-energy limit with $\cos b\rightarrow-1$.

\section{Nonclassical Hamiltonian in the low-energy limit}

In canonical LQC, the effective Hamiltonian is crucial for studying the evolution of the Universe.   To describe the evolution of background and perturbation more clearly, in this section, we aim to find the effective Hamiltonian in the low-energy limit with $\cos b\rightarrow-1$. At the end of this section, we find that this Hamiltonian does exist, and it can be expressed in terms of the Arnowitt-Deser-Misner (ADM) variables or the Ashtekar variables. For the sake of simplicity, we first use the ADM formalism to derive the Hamiltonian.

\subsection{Background Hamiltonian in the low-energy limit}

In the ADM formalism, the Hamiltonian of classical STT is given by the combination of the  Hamiltonian constraint $\mathcal{C}$ smeared on some fiducial cell $\Sigma$ with lapse function $N$ and the smeared diffeomorphism constraint $\mathcal{D}_a$ with the shift vector $N^a$,
\ba
\textbf{H}[N,N^a]=\int_{\Sigma}d^3x\big(N\mathcal{C}+N^a\mathcal{D}_a\big),\label{CHamiltonian}
\ea
where the Hamiltonian constraint is expressed in terms of the ADM variables as \cite{Han:2015}
\ba
 \mathcal{C}&=&\frac{1}{\sqrt{\det(q)}}\Bigg[\frac{2\kappa\big(q_{ac}q_{bd}
 -\frac{1}{2}q_{ab}q_{cd}\big)p^{ab}p^{cd}}{F(\phi)}+\frac{\big(-F'(\phi)q_{ab}p^{ab}+F(\phi)\pi\big)^2}{2F(\phi)G(\phi)}\Bigg]\nn\\
&&+\sqrt{\det(q)}\bigg[-\frac{1}{2\kappa}F(\phi )R^{(3)}+\frac{1}{\kappa}q^{ab}D_a D_b F(\phi )+\frac{K(\phi)}{2}q^{ab}(D_a\phi)D_b\phi+V(\phi)\bigg]\nn\\
&=&0,\label{admconstraint}
\ea
and the diffeomorphism constraint is expressed as
\ba
\mathcal{D}_a=-2q_{ab}D_c p^{bc}+\pi D_a\phi=0,\label{Dcons}
\ea
in which the canonical variables satisfy the elementary Poisson brackets,
\ba
 \big\{q_{ab}(\vec{x}),p^{cd}(\vec{y})\big\}=\delta^c_{(a}\delta^d_{b)}\delta^{(3)}(\vec{x}-\vec{y}),\quad
  \{\phi(\vec{x}),\pi(\vec{y})\}=\delta^{(3)}(\vec{x}-\vec{y}).
\ea

 On the spatially flat FRW background, the line element of the homogenous part of the spacetime metric reads
 \ba
 ds^2=-\bar{N}^2d\tau^2+a^2(\tau)\big(dx^2_1+dx^2_2+dx^2_3\big),\label{FRWmetric}
 \ea
 where $\bar{N}$ is the homogenous part of lapse function and $a^2\delta_{ab}$ is the homogenous part of the spatial metric $q_{ab}$. In the following, we denote the homogenous parts of $p_{ab}$, $\phi$, and $\pi$ by $p\delta^{ab}$, $\bar{\phi}$, and $\bar{\pi}$ respectively. On the background level, the diffeomorphism constraint vanishes and the Hamiltonian is given by
 \ba
 \textbf{H}^{(0)}\big[\bar{N}\big]=\int_{\Sigma}d^3x\bar{N}\mathcal{C}^{(0)},\label{BHFRW}
 \ea
 where
 \ba
 \mathcal{C}^{(0)}=-\frac{3\kappa ap^2}{F\big(\bar{\phi}\big)}
 +\frac{\Big(-F'\big(\bar{\phi}\big)3a^2p
 +F\big(\bar{\phi}\big)\bar{\pi}\Big)^2}{2a^3F\big(\bar{\phi}\big)G\big(\bar{\phi}\big)
 }+a^3V\big(\bar{\phi}\big),\label{C0}
 \ea
 in which the fundamental variables obey the commutation relationship,
\ba
\big\{a^2,p\big\}=\frac{1}{3V_o},\quad\big\{\bar{\phi},\bar{\pi}\}=\frac{1}{V_o},\label{PBADMB}
\ea
where $V_o\equiv\int_{\Sigma}d^3x$. Then, using the Hamilton's equation,
 \ba
 \frac{dO}{d\tau}=\big\{O,\textbf{H}^{0}\big[\bar{N}\big]\big\},
 \ea
 where $O$ is a function of the background variables, it is easy to obtain the classical background equations of motion.

 In the low-energy limit with $\cos b\rightarrow-1$, the effective Friedmann equation and Klein-Gordon equation reduce to
\ba
&&\bigg(F\big(\bar{\phi}\big)H-\frac{1}{2}\dot{F}\big(\bar{\phi}\big)\bigg)^2
=\frac{\kappa}{3}\rho_e,\label{eomFbminus}\\
&&\ddot{\bar{\phi}}+3H\dot{\bar{\phi}}+\frac{1}{2}\frac{\dot{G}\big(\bar{\phi}\big)}{G\big(\bar{\phi}\big)}
\dot{\bar{\phi}}
+\frac{4F'\big(\bar{\phi}\big)V\big(\bar{\phi}\big)+F\big(\bar{\phi}\big)V'\big(\bar{\phi}\big)}{G\big(\bar{\phi}\big)}
=0.\label{eomKbminus}
\ea
We suppose that like the classical background Hamiltonian (\ref{BHFRW}) the above equations of motion can also be obtained from some background effective Hamiltonian expressed in terms of the ADM variables,
 \ba
\textbf{H}^{(0)}_{b_{-}}\big[\bar{N}\big]=\int_{\Sigma}d^3x\bar{N}\mathcal{C}^{(0)}_{b_{-}}.\label{BHFRW2}
 \ea
in which $\mathcal{C}^{(0)}_{b_{-}}$ is the quantum effective background Hamiltonian constraint in the limit $\cos b\rightarrow-1$, which can be regarded as the classical background Hamiltonian density plus the quantum correction function $\mathcal{Q}(a,p,\bar{\phi},\bar{\pi})$,
\ba
\mathcal{C}^{(0)}_{b_{-}}=\mathcal{C}^{(0)}+\mathcal{Q}\big(a,p,\bar{\phi},\bar{\pi}\big)=0.\label{HDFRW2}
\ea
In the Appendix, we show that the function $\mathcal{Q}$ is given by
\ba
\mathcal{Q}\big(a,p,\bar{\phi},\bar{\pi}\big)=\frac{6pF'\big(\bar{\phi}\big)\bar{\pi}}{aG\big(\bar{\phi}\big)}.
\label{qcterm}
\ea
Thus, the background Hamiltonian density in the limit $\cos b\rightarrow-1$ can be expressed in terms of the background ADM variables as
\ba
 \mathcal{C}^{(0)}_{b_{-}}=&&-\frac{3\kappa ap^2}{F\big(\bar{\phi}\big)}
 +\frac{\Big(3F'\big(\bar{\phi}\big)a^2p
 +F\big(\bar{\phi}\big)\bar{\pi}\Big)^2}{2a^3F\big(\bar{\phi}\big)G\big(\bar{\phi}\big)}
 +a^3V\big(\bar{\phi}\big).\label{C0bminus}
\ea
 Comparing (\ref{C0bminus}) with the classical background Hamiltonian density (\ref{C0}), we find that they differ only by the sign of the term $3F'\big(\bar{\phi}\big)a^2p$.

Now, we discuss the relationship between the two background Hamiltonians (\ref{BHFRW}) and (\ref{BHFRW2}) and the background Hamiltonian of LQC of STT \cite{Han:2019},
\ba
\textbf{H}^{(0)}_{LQC}\big[\bar{N}\big]=&&\int_{\Sigma}d^3x\bar{N}
\Bigg[-\frac{3}{\kappa \gamma^2}\frac{a^3}{F\big(\bar{\phi}\big)}\frac{\sin^2 b}{\Delta}+\frac{\Big(3F'\big(\bar{\phi}\big)\frac{a^3\sin b}{\kappa\gamma\sqrt{\Delta}}
 +F\big(\bar{\phi}\big)\bar{\pi}\Big)^2}{2a^3F\big(\bar{\phi}\big)G\big(\bar{\phi}\big)}
 +a^3V\big(\bar{\phi}\big)\Bigg].\label{BHFRWLQC}
\ea

Using the following commutation relation,
\ba
\big\{b,a^3\big\}=\frac{\kappa\gamma\sqrt{\Delta}}{2V_o},
\ea
we find that
\ba
\bigg\{a^2,-\frac{a\sin b}{\kappa\gamma\sqrt{\Delta}}\bigg\}=\frac{\cos b}{3V_o}.\label{PBsinb}
\ea
Comparing (\ref{PBsinb}) with the commutation relation in (\ref{PBADMB}), we find that in the low-energy limit with $\cos b\rightarrow1$ the conjugate momentum $p$ corresponds to $-\frac{a\sin b}{\kappa\gamma\sqrt{\Delta}}$, while in the low-energy limit with $\cos b\rightarrow-1$ the conjugate momentum $p$ corresponds to $\frac{a\sin b}{\kappa\gamma\sqrt{\Delta}}$. Recall that $\sin^2b=\frac{\rho_e}{\rho_c}$, and we conclude that the classical background Hamiltonian (\ref{BHFRW}) and the background Hamiltonian (\ref{BHFRW2}) can be regarded as two different limiting cases of the background Hamiltonian (\ref{BHFRWLQC}) of LQC.
\subsection{Anomaly-free constraints}
To obtain more information relevant to observations in the low-energy limit with $\cos b\rightarrow-1$,  we need to explore the theory beyond the background level. Inspired by the existence of the background Hamiltonian (\ref{BHFRW2}), it is reasonable to assume that on the spatially flat FRW background a more general Hamiltonian exists. We suppose that this Hamiltonian can also be written as a linear combination of the constraints like the classical theory,
\ba
\textbf{H}_{b_{-}}[N,N^a]=\int_{\Sigma}d^3x \big(N \mathcal{C}_{b_{-}}+N^a \mathcal{D}_a\big),\label{Hamb-}
\ea
in which we assume that the diffeomorphism constraint $\mathcal{D}_a$ keeps its classical expression (\ref{Dcons}) because in loop quantum gravity the diffeomorphism constraint does not receive quantum corrections, and only the Hamiltonian constraint does. Inspired by the expressions of the background Hamiltonian constraint (\ref{C0bminus}) and the classical full Hamiltonian constraint (\ref{admconstraint}), we suppose that the Hamiltonian constraint in the low-energy limit with $\cos b\rightarrow-1$ takes the following form in ADM formalism,
\ba
\mathcal{C}_{b_{-}}&=&\frac{1}{\sqrt{\det(q)}}\Bigg[\frac{2\kappa\big(q_{ac}q_{bd}
 -\frac{1}{2}q_{ab}q_{cd}\big)p^{ab}p^{cd}}{F(\phi)}+\frac{\big(F'(\phi)q_{ab}p^{ab}+F(\phi)\pi\big)^2}{2F(\phi)G(\phi)}\Bigg]\nn\\
&&+\sqrt{\det(q)}\bigg[-\frac{1}{2\kappa}f(\phi)R^{(3)}+\frac{1}{\kappa}q^{ab}D_a D_b g(\phi )+\frac{h(\phi)}{2}q^{ab}(D_a\phi)D_b\phi+V(\phi)\bigg]\nn\\
&=&0,\label{admconstraint2}
\ea
in which $f(\phi)$, $g(\phi)$, and $h(\phi)$ are undetermined functions of $\phi$. On the background level, the terms containing the spatial derivatives vanish, and (\ref{admconstraint2}) reduces to the background Hamiltonian constraint (\ref{C0bminus}). Although the forms of $f(\phi)$, $g(\phi)$, and $h(\phi)$ do not affect the background dynamics,  it does not mean that these functions can be arbitrary functions of $\phi$ because they should be subject to some restrictions. To be specific, we require that the effective theory in the low-energy limit with $\cos b\rightarrow-1$ should have a closed constraint algebra, which means the effective theory should be manifestly consistent. When calculating the constraint algebra, the terms that cannot be expressed as functions of the constraints are called anomalies. The consistency condition requires that any anomaly in the constraint algebra should vanish. In LQC with inverse-volume corrections or holonomy corrections, it turns out that this consistency condition is strong enough to determine the structure of the quantum constraint algebra and fix most of the undermined functions in the constraint \cite{Bojowald:2008,Barrau:2015,Cailleteau:2012a,Cailleteau:2012b,Han:2017,Han:2018}.

In the context that we consider in this paper, the diffeomorphism constraint keeps the classical form. Since the Hamiltonian constraint (\ref{admconstraint2}) is a tensor density of weight one, the Poisson bracket between the smeared diffeomorphism constraint and the smeared Hamiltonian constraint is naturally proportional to the Hamiltonian constraint. Thus, we only need to focus on the Poisson bracket between two smeared Hamiltonian constraints.
\begin{widetext}
Denoting $\textbf{C}_{b_{-}}[M]\equiv\int_{\Sigma}d^3x M\mathcal{C}_{b_{-}}$ and
$\textbf{D}[N^a]\equiv\int_{\Sigma}d^3xN^a\mathcal{D}_a$,
we have
\ba
\big\{\textbf{C}_{b_{-}}[M],\textbf{C}_{b_{-}}[N]\big\}
&=&\textbf{D}\bigg[\frac{f(\phi)}{F(\phi)}q^{ab}(MD_bN-ND_bM)\bigg]
+\int_{\Sigma}d^3x q^{ab}\pi(MD_aN-ND_aM)(D_b\phi)\mathcal{A}_{1}\nn\\
&&+\int_{\Sigma}d^3x p^{ab}(MD_aN-ND_aM)(D_b\phi)\mathcal{A}_{2}
+\int_{\Sigma}d^3x (q_{cd}p^{cd})q^{ab}(MD_aN-ND_aM)(D_b\phi)\mathcal{A}_{3}\nn\\
&&+\int_{\Sigma}d^3x (q_{cd}p^{cd})q^{ab}(MD_aD_bN-ND_aD_bM)\mathcal{A}_{4}
+\int_{\Sigma}d^3x \pi q^{ab}(MD_aD_bN-ND_aD_bM)\mathcal{A}_{5},\nn\\
\ea
in which $\mathcal{A}_{1}$ to $\mathcal{A}_{5}$ are anomalies which read explicitly as
\begin{gather*}
\mathcal{A}_{1}=-\frac{f(\phi)}{F(\phi)}-\frac{2}{\kappa}\frac{F'(\phi)f'(\phi)}{ G(\phi)}+\frac{1}{2\kappa}\frac{F'(\phi)g'(\phi)}{ G(\phi)}+\frac{F(\phi)h(\phi)}{G(\phi)},\label{A1}\\
\mathcal{A}_{2}=2\frac{f(\phi)F'(\phi)}{F^2(\phi)}
+2\frac{f'(\phi)}{F(\phi)}-4\frac{g'(\phi)}{F(\phi)},\label{A2}\\
\mathcal{A}_{3}=-\frac{2}{\kappa}\frac{(F'(\phi))^2f'(\phi)}{ F(\phi)G(\phi)}
+\frac{1}{2\kappa}\frac{(F'(\phi))^2g'(\phi)}{F(\phi)G(\phi)}
+\frac{F'(\phi)h(\phi)}{G(\phi)}
+\frac{g'(\phi)}{F(\phi)},\label{A3}\\
\mathcal{A}_{4}=-\frac{1}{\kappa}\frac{(F'(\phi))^2f(\phi)}{F(\phi)G(\phi)}
-\frac{1}{\kappa}\frac{F'(\phi)g'(\phi)}{G(\phi)},\label{A4}\\
\mathcal{A}_{5}=-\frac{1}{\kappa}\frac{F'(\phi)f(\phi)}{G(\phi)}
-\frac{1}{\kappa}\frac{F(\phi)g'(\phi)}{G(\phi)}.\label{A5}
\end{gather*}
\end{widetext}
Since these anomalies should vanish, we have to solve the equations
\ba
\mathcal{A}_i=0\quad(i=1,2...5).\nn
\ea
At first sight, the number of equations is two more than the number of undermined functions; however, it is not difficult to check that only three out of the five equations are independent.  By solving these equations, we can express the undetermined functions in terms of the known functions $F(\phi)$ and $K(\phi)$,
\ba
f(\phi)&=&\frac{B}{F^3(\phi)},\label{fphi}\\
g(\phi)&=&\frac{B}{3}\frac{1}{F^3(\phi)},\label{gphi}\\
h(\phi)&=&B\bigg(\frac{K(\phi)}{F^4(\phi)}-\frac{4}{\kappa}\frac{(F'(\phi))^2}{F(\phi)^5}\bigg),
\ea
where $B$ is an arbitrary constant. Furthermore, by requiring that  the Hamiltonian constraint (\ref{admconstraint2}) should reduce to the Hamiltonian constraint of the minimally coupled case when $F(\phi)=K(\phi)=1$, we obtain $B=1$. Hence, the anomaly-free Hamiltonian constraint in the low-energy limit with $\cos b\rightarrow-1$ reads
\ba
\mathcal{C}_{b_{-}}&=&\frac{1}{\sqrt{\det(q)}}\Bigg[\frac{2\kappa\big(q_{ac}q_{bd}
 -\frac{1}{2}q_{ab}q_{cd}\big)p^{ab}p^{cd}}{F(\phi)}+\frac{\big(F'(\phi)q_{ab}p^{ab}+F(\phi)\pi\big)^2}{2F(\phi)G(\phi)}\Bigg]\nn\\
&&+\sqrt{\det(q)}\Bigg[-\frac{1}{2\kappa}\frac{1}{F^3(\phi)}R^{(3)}
+\frac{1}{3\kappa}q^{ab}D_a D_b \frac{1}{F^3(\phi)}+\frac{1}{2}\bigg(\frac{K(\phi)}{F^4(\phi)}
-\frac{4}{\kappa}\frac{(F'(\phi))^2}{F^5(\phi)}\bigg)q^{ab}(D_a\phi)D_b\phi+V(\phi)\Bigg]\nn\\
&=&0.\label{Hamconstraint2}
\ea
The existence of (\ref{Hamconstraint2}) guarantees that on the perturbation level the covariance of the theory is maintained not only to linear order of perturbation but also up to all higher order perturbations.

The Poisson bracket between two smeared Hamiltonian constraints reads
\ba
\big\{\textbf{C}_{b_{-}}[M],\textbf{C}_{b_{-}}[N]\big\}
=\textbf{D}\bigg[\frac{1}{F^4(\phi)}q^{ab}(MD_bN-ND_bM)\bigg],\label{PoissonHam}
\ea
in which the prefactor $\frac{1}{F^4(\phi)}$ denotes the quantum modification of the constraint algebra.

It should be pointed out that our derivation of the low-energy effective Hamiltonian constraint (\ref{Hamconstraint2}) is tied to the spatially flat FRW background, which is different from the fact the the classical Hamiltonian constraint (\ref{admconstraint}) holds independently of any background metric. So far, it is not clear whether our result remains unchanged on other spacetime backgrounds.

We mention that we choose ADM variables to express the constraints only for convenience of calculation. In fact, the constraints can also be formulated in terms of the Ashtekar  variables in loop quantum gravity. We can extend the ADM phase space to the phase space of connection variables by introducing the su(2)-valued triad $e^a_i$ and its co-triad $e_a^i$ which satisfy
 $q_{ab}=e_a^ie_b^j\delta_{ij}$, $q^{ab}=e^a_ie^b_j\delta^{ij}$. In the new phase space, the basic variables are the densitized triad and its conjugate momentum,
 \ba
 E^a_i\equiv\sqrt{\det(q)}e^a_i,\quad
 K_a^i\equiv\frac{2\kappa}{\sqrt{\det(q)}}\Big[p^{bc}q_{ab}e_c^i
 -\frac{1}{2}\big(p^{bc}q_{bc}\big)e_a^i\Big],
 \ea
 using which we can define the Ashtekar connection
 $A_a^i\equiv\Gamma_a^i+\gamma K_a^i$
  which satisfies
 \ba
 \big\{A_a^i\big(\vec{x}\big),E^b_j\big(\vec{y}\big)\big\}
 =\gamma\kappa\delta_j^i\delta^b_a\delta^{(3)}\big(\vec{x}-\vec{y}\big),
 \ea
 where $\Gamma_a^i$ is the spin connection compatible with the triad.

 The Hamiltonian constraint (\ref{Hamconstraint2}) (modulo the Gauss constraint) can be reexpressed in terms of the Ashtekar variables,
 \begin{widetext}
 \ba
 \mathcal{C}^{(A)}_{b_{-}}&=&\frac{1}{F^3(\phi)}\frac{E^a_iE^b_j}{2\kappa\sqrt{|\det E}|}\Bigg[\epsilon^{ij}_{~~k}F^k_{ab}
 -2\bigg(\gamma^2+F^2(\phi)\bigg)K^i_{[a}K^j_{b]}\Bigg]+\frac{1}{2F(\phi)G(\phi)\sqrt{|\det E|}}
 \bigg[-\frac{1}{\kappa}F'(\phi)K_a^iE^a_i
 +F(\phi)\pi\bigg]^2\nn\\
 &&+\sqrt{|\det E|}\Bigg[-\frac{1}{\kappa}\frac{1}{F^4(\phi)}D^a D_a F(\phi)
 +\frac{1}{2}\bigg(\frac{K(\phi)}{F^4(\phi)}
+\frac{4}{\kappa}\frac{(F'(\phi))^2}{F^5(\phi)}\bigg)(D^a\phi)D_a\phi+V(\phi)\Bigg]\nn\\
&=&0,\label{newconstraint}
 \ea
 where $F^{~~i}_{ab}:=2\partial_{[a}A^i_{b]}+\epsilon^{~~i}_{jk}A^j_aA^k_b$ is the curvature of Ashtekar connection. The diffeomorphism constraint and the Gauss constraint retain their classical form which should be expressed in terms of the Ashtekar variables. As expected, the expression in (\ref{newconstraint}) is more complex than the one in (\ref{Hamconstraint2}). In the following calculation, we still use the ADM variables.
\end{widetext}

\subsection{The Einstein frame formulation}
It is well known that based on different choices of fundamental variables the classical STT can be formulated in the Jordan frame or the Einstein frame. In this subsection, we show that in the low-energy limit with  $\cos b\rightarrow-1$ the constraints in the Jordan frame can also be transformed into the Einstein frame by field redefinitions.

If $F(\phi)>0$, $G(\phi)>0$, we can define
\ba
\tilde{N}\equiv\frac{N}{F^{\frac{5}{2}}(\phi)},&&\quad \tilde{N}^a\equiv N^a,\nn\\
\tilde{q}_{ab}\equiv\frac{q_{ab}}{F(\phi)},&& \quad\tilde{p}^{ab}\equiv F(\phi)p^{ab},\nn\\
\tilde{\phi}\equiv\int d\phi \frac{\sqrt{G(\phi)}}{F(\phi)},&& \quad \tilde{\pi}\equiv\frac{F(\phi)\pi+F'(\phi)q_{ab}p^{ab}}{\sqrt{G(\phi)}}.
\label{canotrans}
\ea
It is easy to check that
\ba
 \big\{\tilde{q}_{ab}(\vec{x}),\tilde{p}^{cd}(\vec{y})\big\}=\delta^c_{(a}\delta^d_{b)}\delta^{(3)}(\vec{x}-\vec{y}),\quad \{\tilde{\phi}(\vec{x}),\tilde{\pi}(\vec{y})\}=\delta^{(3)}(\vec{x}-\vec{y}),
\ea
and all the other Poisson brackets between the above canonical variables are vanishing. It can be directly checked that the Hamiltonian (\ref{Hamb-}) can  be rewritten in terms of the redefined variables as
\ba
\textbf{H}^{(E)}_{b_{-}}\Big[\tilde{N},\tilde{N}^a\Big]=\int_{\Sigma}d^3x \Big(\tilde{N} \mathcal{C}^{(E)}+\tilde{N}^a \mathcal{D}^{(E)}_a\Big),
\ea
where
\ba
\mathcal{C}^{(E)}&=&\frac{1}{\sqrt{\det(\tilde{q})}}\bigg[2\kappa\bigg(\tilde{q}_{ac}\tilde{q}_{bd}
 -\frac{1}{2}\tilde{q}_{ab}\tilde{q}_{cd}\bigg)\tilde{p}^{ab}\tilde{p}^{cd}\bigg]
 -\frac{\sqrt{\det(\tilde{q})}}{2\kappa}\tilde{R}^{(3)}+\frac{\tilde{\pi}^2}{2\sqrt{\det(\tilde{q})}}+\sqrt{\det(\tilde{q})}\bigg[\frac{1}{2}\tilde{q}^{ab}\Big(\tilde{D}_a\tilde{\phi}\Big)\tilde{D}_b\tilde{\phi}
+\tilde{V}\big(\phi\big)\bigg]\nn\\
&=&0,\label{admconstraintE}\\
\mathcal{D}^{(E)}_a&=&-2\tilde{q}_{ab}\tilde{D}_c \tilde{p}^{bc}+\tilde{\pi} \tilde{D}_a\tilde{\phi}=0,
\label{DconstraintE}
\ea
 in which $\tilde{R}^{(3)}$ is the curvature scalar of the rescaled metric $\tilde{q}_{ab}$, $\tilde{q}^{ab}$ is the inverse of the rescaled metric, $\tilde{D}_a$ is the derivative compatible with the rescaled metric, and $\tilde{V}\big(\phi\big)\equiv F^4(\phi)V(\phi)$.

The Hamiltonian constraint (\ref{admconstraintE}) and the diffeomorphism constraint (\ref{DconstraintE}) are exactly of the form of the minimally coupled case. In this sense, we claim that the theory  in the low-energy limit with  $\cos b\rightarrow-1$ can also be transformed into the Einstein frame by field redefinitions. Nevertheless, the frame transformation in this case is different from that in the classical case. In the Hamiltonian formalism of the classical theory, the transformation from the Jordan frame to the Einstein frame is accomplished by the following redefinition of variables,
\ba
\hat{N}\equiv\sqrt{F(\phi)}N,&&\quad \hat{N}^a\equiv N^a,\nn\\
\hat{q}_{ab}\equiv F(\phi)q_{ab},&& \quad \hat{p}^{ab}\equiv\frac{p^{ab}}{F(\phi)},\nn\\
\hat{\phi}\equiv\int d\phi \frac{\sqrt{G(\phi)}}{F(\phi)},&& \quad \hat{\pi}\equiv\frac{F(\phi)\pi-F'(\phi)q_{ab}p^{ab}}{\sqrt{G(\phi)}},
\label{canotrans2}
\ea
in which the variables with a hat denote variables of the Einstein frame. Note that in the Hamiltonian formalism the variables $N^2$ and $q_{ab}$ transform in the same way, which correspond to the spacetime metric redefinition $\hat{g}_{ab}\equiv F(\phi)g_{ab}$ in the Lagrangian formalism of the classical STT. In (\ref{canotrans}), the variables $N^2$ and $q_{ab}$ transform in different ways, which do not correspond to any spacetime metric redefinition. In fact, there does not exist a Lagrangian that can yield the Hamiltonian in Eq. (\ref{Hamb-}) by Legendre transformation, which is not unusual in LQC since there does not exist a manifestly covariant Lagrangian which can yield the equations of motion of canonical LQC either.

It is natural to ask whether the Jordan frame and the Einstein frame are physically equivalent in the limit $\cos b\rightarrow-1$. In Sec. V, we compare the results of the two frames in the case of slow-roll inflation.
\section{Cosmological perturbations}
In this section, we construct the linear perturbation theory on the spatially flat FRW background.  First of all, we split the variables as
\ba
N=\bar{N}+\delta N,&&\quad N^a=\delta N^a,\nn\\
q_{ab}=a^2\delta_{ab}+\delta q_{ab},&&\quad p^{ab}=p\delta^{ab}+\delta p^{ab},\nn\\
\phi=\bar{\phi}+\delta\phi,&&\quad \pi=\bar{\pi}+\delta\pi;
\ea
then, we expand the Hamiltonian (\ref{Hamb-}) to second order of perturbations,
\ba
\textbf{H}_{b_{-}}\big[\bar{N},\delta N,\delta N^a\big]
&=&\textbf{H}_{b_{-}}^{(0)}\big[\bar{N}\big]
+\textbf{H}_{b_{-}}^{(2)}\big[\bar{N}\big]
+\textbf{H}_{b_{-}}^{(2)}\big[\delta N\big]+\textbf{D}^{(2)}\big[\delta N^a\big],\label{Hsecp}
\ea
where
\ba
\textbf{H}_{b_{-}}^{(0)}\big[\bar{N}\big]\equiv\int_{\Sigma}d^3x\bar{N}\mathcal{C}^{(0)}_{b_{-}},\quad
\textbf{H}_{b_{-}}^{(2)}\big[\bar{N}\big]\equiv\int_{\Sigma}d^3x\bar{N}\mathcal{C}^{(2)}_{b_{-}},\quad
\textbf{H}_{b_{-}}^{(2)}\big[\delta N\big]\equiv\int_{\Sigma}d^3x\delta N\mathcal{C}^{(1)}_{b_{-}},\quad
\textbf{D}^{(2)}\big[\delta N\big]\equiv\int_{\Sigma}d^3x\delta N^a\mathcal{D}^{(1)}_{a},\nn\\\label{pcons}
\ea
in which the expression of $\mathcal{C}^{(0)}_{b_{-}}$ is given in Eq. (\ref{C0bminus}), and $\mathcal{D}^{(1)}_{a}$, $\mathcal{C}^{(1)}_{b_{-}}$ denote the linearly perturbed diffeomorphism constraint and the linearly perturbed Hamiltonian constraint respectively,
\begin{widetext}
\ba
\mathcal{D}^{(1)}_{a}&=&-2a^2\delta_{ac}\partial_b\delta p^{bc}-2p\delta^{bc}\partial_c\delta q_{ab}
+p\delta^{bc}\partial_a\delta q_{bc}+\bar{\pi}\partial_a\delta\phi=0,\label{perdc}\\
\mathcal{C}^{(1)}_{b_{-}}&=&\Bigg(-\frac{\kappa p^2}{2aF\big(\bar{\phi}\big)}
+\frac{3p^2\big(F'\big(\bar{\phi}\big)\big)^2}{4aF\big(\bar{\phi}\big)G\big(\bar{\phi}\big)}
-\frac{pF'\big(\bar{\phi}\big)\bar{\pi}}{2a^3G\big(\bar{\phi}\big)}
-\frac{F\big(\bar{\phi}\big)\bar{\pi}^2}{4a^5G\big(\bar{\phi}\big)}
+\frac{a}{2}V\big(\bar{\phi}\big)\Bigg)\delta^{ab}\delta q_{ab}\nn\\
&&+\Bigg(-\frac{2\kappa ap}{F\big(\bar{\phi}\big)}
+\frac{3ap\big(F'\big(\bar{\phi}\big)\big)^2}{F\big(\bar{\phi}\big)G\big(\bar{\phi}\big)}
+\frac{F'\big(\bar{\phi}\big)\bar{\pi}}{aG\big(\bar{\phi}\big)}\Bigg)\delta_{ab}\delta p^{ab}\nn\\
&&+\Bigg[-3\kappa ap^2\Bigg(\frac{1}{F\big(\bar{\phi}\big)}\Bigg)'
+\frac{9}{2}ap^2\Bigg(\frac{\big(F'\big(\bar{\phi}\big)\big)^2}{F\big(\bar{\phi}\big)G\big(\bar{\phi}\big)}\Bigg)'
+\frac{3p\bar{\pi}}{a}\Bigg(\frac{F'\big(\bar{\phi}\big)}{G\big(\bar{\phi}\big)}\Bigg)'
+\frac{\bar{\pi}^2}{2a^3}\Bigg(\frac{F\big(\bar{\phi}\big)}{G\big(\bar{\phi}\big)}\Bigg)'
+a^3V'\big(\bar{\phi}\big)\Bigg]\delta \phi\nn\\
&&+\Bigg(\frac{3pF'\big(\bar{\phi}\big)}{aG\big(\bar{\phi}\big)}
+\frac{F\big(\bar{\phi}\big)\bar{\pi}}{a^3G\big(\bar{\phi}\big)}\Bigg)\delta \pi
+\frac{1}{a\kappa F^3\big(\bar{\phi}\big)}\delta^{a[b}\delta^{c]d}\partial_{c}\partial_{d}\delta q_{ab}
+\frac{a}{3\kappa}
\Bigg(\frac{1}{F^3\big(\bar{\phi}\big)}\Bigg)'\delta^{ab}\partial_{a}\partial_{b}\delta\phi\nn\\
&=&0,\label{perHc}
\ea
and $\mathcal{C}^{(2)}_{b_{-}}$ is given by
\ba
\mathcal{C}^{(2)}_{b_{-}}&=&\Bigg(-\frac{3\kappa p^2}{8a^3F\big(\bar{\phi}\big)}
-\frac{7p^2\big(F'\big(\bar{\phi}\big)\big)^2}{16a^3F\big(\bar{\phi}\big)G\big(\bar{\phi}\big)}
-\frac{pF'\big(\bar{\phi}\big)\bar{\pi}}{8a^5G\big(\bar{\phi}\big)}
+\frac{F\big(\bar{\phi}\big)\bar{\pi}^2}{16a^7G\big(\bar{\phi}\big)}
+\frac{V\big(\bar{\phi}\big)}{8a}\Bigg)\big(\delta^{ab}\delta q_{ab}\big)^2\nn\\
&&+\Bigg(\frac{5\kappa p^2}{4a^3F\big(\bar{\phi}\big)}
+\frac{9p^2\big(F'\big(\bar{\phi}\big)\big)^2}{8a^3F\big(\bar{\phi}\big)G\big(\bar{\phi}\big)}
+\frac{3pF'\big(\bar{\phi}\big)\bar{\pi}}{4a^5G\big(\bar{\phi}\big)}
+\frac{F\big(\bar{\phi}\big)\bar{\pi}^2}{8a^7G\big(\bar{\phi}\big)}
-\frac{V\big(\bar{\phi}\big)}{4a}\Bigg)\delta^{ac}\delta^{bd}\delta q_{ab}\delta q_{cd}\nn\\
&&+\Bigg(-\frac{\kappa a}{F\big(\bar{\phi}\big)}
+\frac{a\big(F'\big(\bar{\phi}\big)\big)^2}{2F\big(\bar{\phi}\big)G\big(\bar{\phi}\big)}\Bigg)
\big(\delta_{ab}\delta p^{ab}\big)^2
+\frac{2\kappa a}{F\big(\bar{\phi}\big)}\delta_{ac}\delta_{bd}\delta p^{ab}\delta p^{cd}
+\frac{F\big(\bar{\phi}\big)}{2a^3G\big(\bar{\phi}\big)}\big(\delta \pi\big)^2\nn\\
&&+\Bigg[-\frac{3\kappa ap^2}{2}\Bigg(\frac{1}{F\big(\bar{\phi}\big)}\Bigg)''
+\frac{9ap^2}{4}\Bigg(\frac{\big(F'\big(\bar{\phi}\big)\big)^2}{F\big(\bar{\phi}\big)G\big(\bar{\phi}\big)}\Bigg)''
+\frac{3a^2p\bar{\pi}}{2}\Bigg(\frac{F'\big(\bar{\phi}\big)}{G\big(\bar{\phi}\big)}\Bigg)''
+\frac{\bar{\pi}^2}{4a^3}\Bigg(\frac{F\big(\bar{\phi}\big)}{G\big(\bar{\phi}\big)}\Bigg)''
+\frac{a^3}{2}V''\big(\bar{\phi}\big)\Bigg]\big(\delta \phi\big)^2\nn\\
&&+\Bigg(\frac{2\kappa p}{aF\big(\bar{\phi}\big)}
+\frac{3p\big(F'\big(\bar{\phi}\big)\big)^2}{aF\big(\bar{\phi}\big)G\big(\bar{\phi}\big)}
+\frac{F'\big(\bar{\phi}\big)\bar{\pi}}{a^3G\big(\bar{\phi}\big)}\Bigg)\delta q_{ab}\delta p^{ab}
-\Bigg(\frac{\kappa p}{aF\big(\bar{\phi}\big)}
+\frac{3p\big(F'\big(\bar{\phi}\big)\big)^2}{2aF\big(\bar{\phi}\big)G\big(\bar{\phi}\big)}
+\frac{F'\big(\bar{\phi}\big)\bar{\pi}}{2a^3G\big(\bar{\phi}\big)}\Bigg)
\big(\delta^{ab}\delta q_{ab}\big)\big(\delta_{cd}\delta p^{cd}\big)\nn\\
&&+\Bigg[-\frac{\kappa p^2}{2a}\Bigg(\frac{1}{F\big(\bar{\phi}\big)}\Bigg)'
+\frac{3p^2}{4a}\Bigg(\frac{\big(F'\big(\bar{\phi}\big)\big)^2}{F\big(\bar{\phi}\big)G\big(\bar{\phi}\big)}\Bigg)'
-\frac{p\bar{\pi}}{2a^3}\Bigg(\frac{F'\big(\bar{\phi}\big)}{G\big(\bar{\phi}\big)}\Bigg)'
-\frac{\bar{\pi}^2}{4a^5}\Bigg(\frac{F\big(\bar{\phi}\big)}{G\big(\bar{\phi}\big)}\Bigg)'
+\frac{a}{2}V'\big(\bar{\phi}\big)\Bigg]\delta^{ab}\delta q_{ab}\delta \phi\nn\\
&&+\Bigg[-2\kappa ap\Bigg(\frac{1}{F\big(\bar{\phi}\big)}\Bigg)'
+3ap\Bigg(\frac{\big(F'\big(\bar{\phi}\big)\big)^2}{F\big(\bar{\phi}\big)G\big(\bar{\phi}\big)}\Bigg)'
+\frac{\bar{\pi}}{a}\Bigg(\frac{F'\big(\bar{\phi}\big)}{G\big(\bar{\phi}\big)}\Bigg)'
\Bigg]\delta_{ab}\delta p^{ab}\delta \phi\nn\\
&&+\Bigg(-\frac{F\big(\bar{\phi}\big)\bar{\pi}}{2a^5G\big(\bar{\phi}\big)}
+\frac{pF'\big(\bar{\phi}\big)}{a^3G\big(\bar{\phi}\big)}\Bigg)\delta^{ab}\delta q_{ab}\delta \pi
+\frac{F'\big(\bar{\phi}\big)}{aG\big(\bar{\phi}\big)}\delta_{ab}\delta p^{ab}\delta \pi
+\Bigg[\frac{3p}{a}\Bigg(\frac{F'\big(\bar{\phi}\big)}{G\big(\bar{\phi}\big)}\Bigg)'
+\frac{\bar{\pi}}{a^3}\Bigg(\frac{F\big(\bar{\phi}\big)}{G\big(\bar{\phi}\big)}\Bigg)'\Bigg]\delta \phi\delta \pi\nn\\
&&+\frac{1}{\kappa a}\Bigg(\frac{1}{F^3\big(\bar{\phi}\big)}\Bigg)'
\delta^{a[b}\delta^{c]d}\big(\partial_c\partial_d\delta q_{ab}\big)\delta \phi
-\frac{1}{4\kappa a^3F^3\big(\bar{\phi}\big)}\big[2\delta^{a[c}\delta^{e]d}\delta^{bf}
+\delta^{a[b}\delta^{c]d}\delta^{ef}\big]\big(\partial_e\delta q_{ab}\big)\partial_f\delta q_{cd}\nn\\
&&+\frac{a}{2}\bigg(\frac{K(\phi)}{F^4(\phi)}
-\frac{4}{\kappa}\frac{(F'(\phi))^2}{F^5(\phi)}\bigg)\delta^{ab}\big(\partial_a\delta\phi\big)\partial_b\delta\phi.\label{pHam}
\ea

Using the Hamiltonian (\ref{Hsecp}), we can derive the Hamilton's equations of motion of the perturbed variables,
\ba
\frac{d \delta q_{ab}}{d\tau}&=&2a^2\delta_{c(a}\partial_{b)}\delta N^{c}
+\Bigg(-\frac{2\kappa ap}{F\big(\bar{\phi}\big)}
+\frac{3ap\big(F'\big(\bar{\phi}\big)\big)^2}{F\big(\bar{\phi}\big)G\big(\bar{\phi}\big)}
+\frac{F'\big(\bar{\phi}\big)\bar{\pi}}{aG\big(\bar{\phi}\big)}\Bigg)\delta N\delta_{ab}\nn\\
&&+\bar{N}\Bigg[
\Bigg(-\frac{2\kappa a}{F\big(\bar{\phi}\big)}
+\frac{a\big(F'\big(\bar{\phi}\big)\big)^2}{F\big(\bar{\phi}\big)G\big(\bar{\phi}\big)}\Bigg)
\big(\delta_{cd}\delta p^{cd}\big)\delta_{ab}+\frac{4\kappa a}{F\big(\bar{\phi}\big)}\delta_{ac}\delta_{bd}\delta p^{cd}+\Bigg(\frac{2\kappa p}{aF\big(\bar{\phi}\big)}
+\frac{3p\big(F'\big(\bar{\phi}\big)\big)^2}{aF\big(\bar{\phi}\big)G\big(\bar{\phi}\big)}
+\frac{F'\big(\bar{\phi}\big)\bar{\pi}}{a^3G\big(\bar{\phi}\big)}\Bigg)\delta q_{ab}\nn\\
&&-\Bigg(\frac{\kappa p}{aF\big(\bar{\phi}\big)}
+\frac{3p\big(F'\big(\bar{\phi}\big)\big)^2}{2aF\big(\bar{\phi}\big)G\big(\bar{\phi}\big)}
+\frac{F'\big(\bar{\phi}\big)\bar{\pi}}{2a^3G\big(\bar{\phi}\big)}\Bigg)
\big(\delta^{cd}\delta q_{cd}\big)\delta_{ab}
+\frac{F'\big(\bar{\phi}\big)}{aG\big(\bar{\phi}\big)}\delta \pi\delta_{ab}\nn\\
&&+\Bigg(-2\kappa ap\Bigg(\frac{1}{F\big(\bar{\phi}\big)}\Bigg)'
+3ap\Bigg(\frac{\big(F'\big(\bar{\phi}\big)\big)^2}{F\big(\bar{\phi}\big)G\big(\bar{\phi}\big)}\Bigg)'
+\frac{\bar{\pi}}{a}\Bigg(\frac{F'\big(\bar{\phi}\big)}{G\big(\bar{\phi}\big)}\Bigg)'
\Bigg)\delta \phi\delta_{ab}\Bigg],\label{ddqdtau}\\
\frac{d \delta p^{ab}}{d\tau}&=&-2p\delta^{c(a}\partial_c\delta N^{b)}
+p\delta^{ab}\partial_c\delta N^{c}
+\Bigg(\frac{\kappa p^2}{2aF\big(\bar{\phi}\big)}
-\frac{3p^2\big(F'\big(\bar{\phi}\big)\big)^2}{4aF\big(\bar{\phi}\big)G\big(\bar{\phi}\big)}
+\frac{pF'\big(\bar{\phi}\big)\bar{\pi}}{2a^3G\big(\bar{\phi}\big)}
+\frac{F\big(\bar{\phi}\big)\bar{\pi}^2}{4a^5G\big(\bar{\phi}\big)}
-\frac{a}{2}V\big(\bar{\phi}\big)\Bigg)\delta N\delta^{ab}\nn\\
&&-\frac{1}{a\kappa F^3\big(\bar{\phi}\big)}\delta^{a[b}\delta^{c]d}\partial_{c}\partial_{d}\delta N
+\bar{N}\Bigg[\Bigg(\frac{3\kappa p^2}{4a^3F\big(\bar{\phi}\big)}
+\frac{7p^2\big(F'\big(\bar{\phi}\big)\big)^2}{8a^3F\big(\bar{\phi}\big)G\big(\bar{\phi}\big)}
+\frac{pF'\big(\bar{\phi}\big)\bar{\pi}}{4a^5G\big(\bar{\phi}\big)}
-\frac{F\big(\bar{\phi}\big)\bar{\pi}^2}{8a^7G\big(\bar{\phi}\big)}
-\frac{V\big(\bar{\phi}\big)}{4a}\Bigg)\big(\delta^{cd}\delta q_{cd}\big)\delta^{ab}\nn\\
&&-\Bigg(\frac{5\kappa p^2}{2a^3F\big(\bar{\phi}\big)}
+\frac{9p^2\big(F'\big(\bar{\phi}\big)\big)^2}{4a^3F\big(\bar{\phi}\big)G\big(\bar{\phi}\big)}
+\frac{3pF'\big(\bar{\phi}\big)\bar{\pi}}{2a^5G\big(\bar{\phi}\big)}
+\frac{F\big(\bar{\phi}\big)\bar{\pi}^2}{4a^7G\big(\bar{\phi}\big)}
-\frac{V\big(\bar{\phi}\big)}{2a}\Bigg)\delta^{ac}\delta^{bd}\delta q_{cd}\nn\\
&&-\Bigg(\frac{2\kappa p}{aF\big(\bar{\phi}\big)}
+\frac{3p\big(F'\big(\bar{\phi}\big)\big)^2}{aF\big(\bar{\phi}\big)G\big(\bar{\phi}\big)}
+\frac{F'\big(\bar{\phi}\big)\bar{\pi}}{a^3G\big(\bar{\phi}\big)}\Bigg)\delta p^{ab}
+\Bigg(\frac{\kappa p}{aF\big(\bar{\phi}\big)}
+\frac{3p\big(F'\big(\bar{\phi}\big)\big)^2}{2aF\big(\bar{\phi}\big)G\big(\bar{\phi}\big)}
+\frac{F'\big(\bar{\phi}\big)\bar{\pi}}{2a^3G\big(\bar{\phi}\big)}\Bigg)
\big(\delta_{cd}\delta p^{cd}\big)\delta^{ab}\nn\\
&&+\Bigg(\frac{\kappa p^2}{2a}\Bigg(\frac{1}{F\big(\bar{\phi}\big)}\Bigg)'
-\frac{3p^2}{4a}\Bigg(\frac{\big(F'\big(\bar{\phi}\big)\big)^2}{F\big(\bar{\phi}\big)G\big(\bar{\phi}\big)}\Bigg)'
+\frac{p\bar{\pi}}{2a^3}\Bigg(\frac{F'\big(\bar{\phi}\big)}{G\big(\bar{\phi}\big)}\Bigg)'
+\frac{\bar{\pi}^2}{4a^5}\Bigg(\frac{F\big(\bar{\phi}\big)}{G\big(\bar{\phi}\big)}\Bigg)'
-\frac{a}{2}V'\big(\bar{\phi}\big)\Bigg)\delta \phi\delta^{ab}\nn\\
&&+\Bigg(\frac{F\big(\bar{\phi}\big)\bar{\pi}}{2a^5G\big(\bar{\phi}\big)}
-\frac{pF'\big(\bar{\phi}\big)}{a^3G\big(\bar{\phi}\big)}\Bigg)\delta \pi\delta^{ab}
-\frac{1}{\kappa a}\Bigg(\frac{1}{F^3\big(\bar{\phi}\big)}\Bigg)'
\delta^{a[b}\delta^{c]d}\partial_c\partial_d\delta \phi\nn\\
&&-\frac{1}{2\kappa a^3F^3\big(\bar{\phi}\big)}\big[2\delta^{a[c}\delta^{e]d}\delta^{bf}
+\delta^{a[b}\delta^{c]d}\delta^{ef}\big]\partial_e\partial_f\delta q_{cd}\Bigg],\label{ddpdtau}\\
\frac{d \delta \phi}{d\tau}&=&\Bigg(\frac{3pF'\big(\bar{\phi}\big)}{aG\big(\bar{\phi}\big)}
+\frac{F\big(\bar{\phi}\big)\bar{\pi}}{a^3G\big(\bar{\phi}\big)}\Bigg)\delta N
+\bar{N}\Bigg[\frac{F\big(\bar{\phi}\big)}{a^3G\big(\bar{\phi}\big)}\delta \pi
+\Bigg(-\frac{F\big(\bar{\phi}\big)\bar{\pi}}{2a^5G\big(\bar{\phi}\big)}
+\frac{pF'\big(\bar{\phi}\big)}{a^3G\big(\bar{\phi}\big)}\Bigg)\delta^{ab}\delta q_{ab}
+\frac{F'\big(\bar{\phi}\big)}{aG\big(\bar{\phi}\big)}\delta_{ab}\delta p^{ab}\nn\\
&&+\Bigg(\frac{3p}{a}\Bigg(\frac{F'\big(\bar{\phi}\big)}{G\big(\bar{\phi}\big)}\Bigg)'
+\frac{\bar{\pi}}{a^3}\Bigg(\frac{F\big(\bar{\phi}\big)}{G\big(\bar{\phi}\big)}\Bigg)'\Bigg)\delta \phi
\Bigg],\label{ddphidtau}\\
\frac{d \delta \pi}{d\tau}&=&\bar{\pi}\partial_a\delta N^a
-\frac{a}{3\kappa}\Bigg(\frac{1}{F^3\big(\bar{\phi}\big)}\Bigg)'\delta^{ab}\partial_{a}\partial_{b}\delta N\nn\\
&&+\Bigg[3\kappa ap^2\Bigg(\frac{1}{F\big(\bar{\phi}\big)}\Bigg)'
-\frac{9}{2}ap^2\Bigg(\frac{\big(F'\big(\bar{\phi}\big)\big)^2}{F\big(\bar{\phi}\big)G\big(\bar{\phi}\big)}\Bigg)'
-\frac{3p\bar{\pi}}{a}\Bigg(\frac{F'\big(\bar{\phi}\big)}{G\big(\bar{\phi}\big)}\Bigg)'
-\frac{\bar{\pi}^2}{2a^3}\Bigg(\frac{F\big(\bar{\phi}\big)}{G\big(\bar{\phi}\big)}\Bigg)'
-a^3V'\big(\bar{\phi}\big)\Bigg]\delta N\nn\\
&&+\bar{N}\Bigg[\Bigg(3\kappa ap^2\Bigg(\frac{1}{F\big(\bar{\phi}\big)}\Bigg)''
-\frac{9ap^2}{2}\Bigg(\frac{\big(F'\big(\bar{\phi}\big)\big)^2}{F\big(\bar{\phi}\big)G\big(\bar{\phi}\big)}\Bigg)''
-3a^2p\bar{\pi}\Bigg(\frac{F'\big(\bar{\phi}\big)}{G\big(\bar{\phi}\big)}\Bigg)''
-\frac{\bar{\pi}^2}{2a^3}\Bigg(\frac{F\big(\bar{\phi}\big)}{G\big(\bar{\phi}\big)}\Bigg)''
-a^3V''\big(\bar{\phi}\big)\Bigg)\delta \phi\nn\\
&&
+\Bigg(\frac{\kappa p^2}{2a}\Bigg(\frac{1}{F\big(\bar{\phi}\big)}\Bigg)'
-\frac{3p^2}{4a}\Bigg(\frac{\big(F'\big(\bar{\phi}\big)\big)^2}{F\big(\bar{\phi}\big)G\big(\bar{\phi}\big)}\Bigg)'
+\frac{p\bar{\pi}}{2a^3}\Bigg(\frac{F'\big(\bar{\phi}\big)}{G\big(\bar{\phi}\big)}\Bigg)'
+\frac{\bar{\pi}^2}{4a^5}\Bigg(\frac{F\big(\bar{\phi}\big)}{G\big(\bar{\phi}\big)}\Bigg)'
-\frac{a}{2}V'\big(\bar{\phi}\big)\Bigg)\delta^{ab}\delta q_{ab}\nn\\
&&+\Bigg(2\kappa ap\Bigg(\frac{1}{F\big(\bar{\phi}\big)}\Bigg)'
-3ap\Bigg(\frac{\big(F'\big(\bar{\phi}\big)\big)^2}{F\big(\bar{\phi}\big)G\big(\bar{\phi}\big)}\Bigg)'
-\frac{\bar{\pi}}{a}\Bigg(\frac{F'\big(\bar{\phi}\big)}{G\big(\bar{\phi}\big)}\Bigg)'
\Bigg)\delta_{ab}\delta p^{ab}
-\Bigg(\frac{3p}{a}\Bigg(\frac{F'\big(\bar{\phi}\big)}{G\big(\bar{\phi}\big)}\Bigg)'
+\frac{\bar{\pi}}{a^3}\Bigg(\frac{F\big(\bar{\phi}\big)}{G\big(\bar{\phi}\big)}\Bigg)'\Bigg)\delta \pi\nn\\
&&-\frac{1}{\kappa a}\Bigg(\frac{1}{F^3\big(\bar{\phi}\big)}\Bigg)'
\delta^{a[b}\delta^{c]d}\big(\partial_c\partial_d\delta q_{ab}\big)
+a\bigg(\frac{K(\phi)}{F^4(\phi)}
-\frac{4}{\kappa}\frac{(F'(\phi))^2}{F^5(\phi)}\bigg)\delta^{ab}\partial_a\partial_b\delta\phi\Bigg].\label{ddpidtau}
\ea
\end{widetext}
\subsection{Gauge invariant variables}
In the canonical theory, the gauge transformation of the perturbed variable is governed by the perturbed diffeomorphism constraint and the perturbed Hamiltonian constraint. Since in our case the perturbed Hamiltonian constraint receives quantum corrections, the gauge transformation of the perturbed variable is also subject to quantum corrections. In this subsection, we construct the gauge invariant variables following the techniques introduced in Ref. \cite{Bojowald:2009}.

 If the
lapse function and the shift vector had local infinitesimal variations,
\ba
N\rightarrow N+\delta v,\quad N^a\rightarrow N^a+\delta v^a,
\ea
the gauge transformations of a perturbed phase space variable $\delta X$ generated by the perturbed Hamiltonian and diffeomorphism constraints are given by
\ba
\delta_{[\delta v,\delta v^a]}\delta X=\big\{\delta X,\textbf{H}_{b_{-}}^{(2)}\big[\delta v\big]+\textbf{D}^{(2)}\big[\delta v^a\big]\big\},\label{gtpv}
\ea
in which the left-hand side of Eq. (\ref{gtpv}) denotes the gauge transformations of  $\delta X$. Using Eq. (\ref{PoissonHam}), it is not difficult to prove that the gauge transformation of the time derivative of a perturbed phase space variable satisfies
\ba
&&\delta_{[\delta v,\delta v^a]}\bigg(\frac{d\delta X}{d\tau}\bigg)
-\frac{d}{d\tau}\bigg(\delta_{[\delta v,\delta v^a]}\delta X\bigg)=\bigg\{\delta X,\textbf{D}^{(2)}\bigg[\frac{\bar{N}}{a^2F^4\big(\bar{\phi}\big)}
\delta^{ab}\partial_b\delta v\bigg]\bigg\}.\label{trddeltaX}
\ea

 Using Eq. (\ref{gtpv}), we derive the gauge transformations of the following perturbed phase space variables:
\ba
\delta_{[\delta v,\delta v^a]}\delta q_{ab}&=&2a^2\delta_{c(a}\partial_{b)}\delta v^{c}
+\Bigg(-\frac{2\kappa ap}{F\big(\bar{\phi}\big)}
+\frac{3ap\big(F'\big(\bar{\phi}\big)\big)^2}{F\big(\bar{\phi}\big)G\big(\bar{\phi}\big)}
+\frac{F'\big(\bar{\phi}\big)\bar{\pi}}{aG\big(\bar{\phi}\big)}\Bigg)\delta v\delta_{ab},\label{gtq}\\
\delta_{[\delta v,\delta v^a]}\delta p^{ab}&=&-2p\delta^{c(a}\partial_c\delta v^{b)}
+p\delta^{ab}\partial_c\delta v^{c}
+\Bigg(\frac{\kappa p^2}{2aF\big(\bar{\phi}\big)}
-\frac{3p^2\big(F'\big(\bar{\phi}\big)\big)^2}{4aF\big(\bar{\phi}\big)G\big(\bar{\phi}\big)}
+\frac{pF'\big(\bar{\phi}\big)\bar{\pi}}{2a^3G\big(\bar{\phi}\big)}
+\frac{F\big(\bar{\phi}\big)\bar{\pi}^2}{4a^5G\big(\bar{\phi}\big)}
-\frac{a}{2}V\big(\bar{\phi}\big)\Bigg)\delta v\delta^{ab},\nn\\\label{gtp}\\
\delta_{[\delta v,\delta v^a]}\delta \phi&=&\Bigg(\frac{3pF'\big(\bar{\phi}\big)}{aG\big(\bar{\phi}\big)}
+\frac{F\big(\bar{\phi}\big)\bar{\pi}}{a^3G\big(\bar{\phi}\big)}\Bigg)\delta v,\label{gtpv2}
\ea

To simplify the analysis, we separately consider different modes of perturbations. For the scalar mode of perturbations, we denote
\ba
\delta N=\bar{N}\varphi,\quad\delta N^a=\delta^{ab}\partial_bB,\quad
\delta q_{ab}=2a^2[-\psi\delta_{ab}+\partial_a\partial_bE].\label{smode}
\ea

From Eq. (\ref{ddqdtau}), we find
\ba
\delta p^{ab}&=&\frac{F\big(\bar{\phi}\big)a}{2\kappa \bar{N}}\Bigg[2\bigg(\frac{1}{a}\frac{da}{d\tau}
-\frac{1}{2}\frac{F'\big(\bar{\phi}\big)}{F\big(\bar{\phi}\big)}\frac{d\bar{\phi}}{d\tau}\bigg)
\bigg[\bigg(\varphi-\psi-\frac{F'\big(\bar{\phi}\big)}{F\big(\bar{\phi}\big)}\delta\phi\bigg)\delta^{ab}
+2\delta^{ac}\delta^{bd}\partial_c\partial_dE-(\delta^{cd}\partial_c\partial_dE)\delta^{ab}\bigg]\nn\\
&&-2\bigg(\frac{d\psi}{d\tau}
-\frac{1}{2}\frac{F'\big(\bar{\phi}\big)}{F\big(\bar{\phi}\big)}\frac{d\delta \phi}{d\tau}\bigg)\delta^{ab}+\delta^{ac}\delta^{bd}\partial_c\partial_d\frac{dE}{d\tau}
-\bigg(\delta^{cd}\partial_c\partial_d\frac{dE}{d\tau}\bigg)\delta^{ab}
-\delta^{ac}\delta^{bd}\partial_c\partial_dB+(\delta^{cd}\partial_c\partial_dB)\delta^{ab}\Bigg].\label{deltap}
\ea

We parametrize the scalar components of the variations by two scalar functions $v_{0}$ and $v$ such that
\ba
\delta v=\bar{N}v_{0},\quad \delta v^{a}=\delta^{ab}\partial_b v.
\ea
In the following, we denote the scalar component of the gauge transformations of $\delta X$ by
$\delta_{[\delta v,\delta v^a]}\delta X=\delta_{[v_0,v]}\delta X$.
  In accord with the standard treatment in classical cosmology, we set $\bar{N}=a$ and denote the corresponding conformal time as $d\eta$. Substituting Eq. (\ref{smode}) into Eqs. (\ref{gtq})~(\ref{gtpv2}), and using Eq. (\ref{trddeltaX}) along with the canonical background equations of motion in (\ref{Hameq}),  we find
\ba
\delta_{[v_0,v]}\psi&=&-\mathcal{H}v_0,\quad \delta_{[v_0,v]} E=v,\quad \delta_{[v_0,v]}\delta\phi=\bar{\phi}_{,\eta}v_0,\quad
\delta_{[v_0,v]}\big(\bar{\psi}_{,\eta}\big)=\big(\delta_{[v_0,v]}\psi\big)_{,\eta},\quad
\delta_{[v_0,v]} \big(E_{,\eta}\big)
=v_{,\eta}+\frac{v_0}{F^4\big(\bar{\phi}\big)},\nn\\\label{trsqab}
\ea
where the subscript ``${\eta}$"  denotes the derivative with respect to the conformal time, i.e.,  ${v}_{,\eta}\equiv\frac{dv}{d\eta}$, and the conformal Hubble parameter $\mathcal{H}$ is defined by $\mathcal{H}\equiv\frac{a_{,\eta}}{a}$.
Substituting Eq. (\ref{deltap}) into Eqs. (\ref{gtq})~(\ref{gtpv2})and using Eq. (\ref{trsqab}), we get
\ba
\delta_{[v_0,v]}\varphi=(v_0)_{,\eta}+\mathcal{H}v_0,\quad
\delta_{[v_0,v]}B=F^2\big(\bar{\phi}\big)v_{,\eta}.\label{trspab}
\ea
From Eqs. (\ref{trsqab}) and (\ref{trspab}), we obtain
\ba
\delta_{[v_0,v]}
\bigg[F^2\big(\bar{\phi}\big)\bigg(B-F^2\big(\bar{\phi}\big)E_{,\eta}\bigg)\bigg]
=-v_0,\quad
\delta_{[v_0,v]}
\Bigg[\bigg(F^2\big(\bar{\phi}\big)\bigg(B-F^2\big(\bar{\phi}\big)E_{,\eta}\bigg)\bigg)_{,\eta}\Bigg]
=-(v_0)_{,\eta}.
\label{dv0vdeta}
\ea
We define the following variables
\ba
&&\Phi\equiv\varphi-\frac{5}{2}\frac{F'\big(\bar{\phi}\big)}{F\big(\bar{\phi}\big)}\delta \phi
+\mathcal{G}\bigg(B-F^2\big(\bar{\phi}\big)E_{,\eta}\bigg)
+F^2\big(\bar{\phi}\big)\Big(B-F^2\big(\bar{\phi}\big)E_{,\eta}\Big)_{,\eta},\label{defPhi}\\
&&\Psi\equiv\psi-\frac{1}{2}\frac{F'\big(\bar{\phi}\big)}{F\big(\bar{\phi}\big)}\delta \phi
-\mathcal{G}
\bigg(B-F^2\big(\bar{\phi}\big)E_{,\eta}\bigg),\label{defPsi}\\
&&\delta\phi^{GI}\equiv\delta\phi+F^2\big(\bar{\phi}\big)\bar{\phi}_{,\eta}
\bigg[\bigg(B-F^2\big(\bar{\phi}\big)E_{,\eta}\bigg)\bigg],\label{defdeltaphiGI}
\ea
where
\ba
\mathcal{G}\equiv F^2\big(\bar{\phi}\big)\bigg(\mathcal{H}
-\frac{F'(\bar{\phi})\bar{\phi}_{,\eta}}{2F(\bar{\phi})}\bigg). \label{defmathG}
\ea
Obviously, $\Phi$ and $\Psi$ can reproduce the Bardeen potentials if $F\big(\bar{\phi}\big)=1$.
Using Eqs. (\ref{trsqab})~(\ref{dv0vdeta}), it is direct to check that
$\delta_{[v_0,v]}\Phi=\delta_{[v_0,v]}\Psi=\delta_{[v_0,v]}\delta\phi^{GI}=0$,
which means these variables are gauge invariant.

\subsection{Evolution equations of gauge invariant variables}
In this subsection, we derive the second order evolution equations of the gauge invariant variables.

 We consider the gauge invariant scalar modes first.  The perturbed constraint equations $\mathcal{D}^{(1)}_{a}=0$ and $\mathcal{C}^{(1)}_{b_{-}}=0$ can be, respectively, rewritten in terms of the gauge invariant variables as
\ba
\partial_a\bigg[F^2\big(\bar{\phi}\big)\Psi_{,\eta}
+\mathcal{G}\Phi\bigg]
&=&\frac{\kappa}{2}G\big(\bar{\phi}\big)\bar{\phi}_{,\eta}
\partial_a\delta \phi^{GI},\label{pdcgi}\\
\nabla^2\Psi
-3F^2\big(\bar{\phi}\big)\mathcal{G}\Psi_{,\eta}
-\bigg[3\mathcal{G}^2
-\frac{\kappa}{2}F^2\big(\bar{\phi}\big)G\big(\bar{\phi}\big)\big(\bar{\phi}_{,\eta}\big)^2\bigg]\Phi
&=&\frac{\kappa}{2}F^2\big(\bar{\phi}\big)
\bigg[G\big(\bar{\phi}\big)\bar{\phi}_{,\eta}\delta\phi^{GI}_{,\eta}
+\Big(4a^2F'\big(\bar{\phi}\big)V\big(\bar{\phi}\big)\nn\\
&&
+a^2F\big(\bar{\phi}\big)V'\big(\bar{\phi}\big)\Big)\delta\phi^{GI}\bigg],
\label{pdHgi}
\ea
where $\nabla^2\equiv\delta^{ab}\partial_a\partial_b$.

Then, substituting Eq. (\ref{deltap}) into Eq. (\ref{ddpdtau}) and taking into account the background equations of motion, it is straightforward to show that the off-diagonal part of Eq. (\ref{ddpdtau}) yields
\ba
\Phi=\Psi,\label{odp}
\ea
and the diagonal part of Eq. (\ref{ddpdtau}) yields the following equation:
\ba
&&F^2\big(\bar{\phi}\big)\Psi_{,\eta\eta}
+2F^2\big(\bar{\phi}\big)\bigg(\mathcal{H}
+\frac{F'\big(\bar{\phi}\big)}{2F\big(\bar{\phi}\big)}\bar{\phi}_{,\eta}\bigg)
\Psi_{,\eta}+\mathcal{G}\Phi_{,\eta}
+\Bigg[\frac{d\mathcal{G}}{d\eta}
+\frac{\mathcal{G}^2}{F^2\big(\bar{\phi}\big)}\Bigg]\Phi\nn\\
&=&\frac{\kappa}{2}
\bigg[G\big(\bar{\phi}\big)\bar{\phi}_{,\eta}\delta\phi^{GI}_{,\eta}
-\Big(4a^2F'\big(\bar{\phi}\big)V\big(\bar{\phi}\big)
+a^2F\big(\bar{\phi}\big)V'\big(\bar{\phi}\big)\Big)\delta\phi^{GI}\bigg],\label{dp}
\ea
where the subscript ``$\eta\eta$" denotes the second derivative with respect to the conformal time.

Moreover, substituting Eq. (\ref{ddphidtau}) into Eq. (\ref{ddpidtau}), we find that the perturbed Klein-Gordon equation can be expressed in terms of the gauge invariant variables,
\ba
&&\delta\phi^{GI}_{,\eta\eta}
+\bigg(\frac{2\mathcal{G}}{F^2\big(\bar{\phi}\big)}
+\frac{1}{G\big(\bar{\phi}\big)}\Big(G\big(\bar{\phi}\big)\Big)_{,\eta}\bigg)\delta\phi^{GI}_{,\eta}
+\mathcal{A}a^2\delta \phi^{GI}-\frac{1}{F^4\big(\bar{\phi}\big)}\nabla^2\delta \phi^{GI}
\nn\\
&&-\bar{\phi}_{,\eta}\bigg[\Phi_{,\eta}+3\Psi_{,\eta}\bigg]
-2\frac{a^2}{G\big(\bar{\phi}\big)}\Big(4F'\big(\bar{\phi}\big)V\big(\bar{\phi}\big)
+F\big(\bar{\phi}\big)V'\big(\bar{\phi}\big)\Big)\Phi=0,\label{d2deltaphi}
\ea
where
\ba
\mathcal{A}&=&-\frac{1}{F\big(\bar{\phi}\big)}\Big(F\big(\bar{\phi}\big)\Big)_{,\eta\eta}
-\frac{1}{4}\frac{1}{G^2\big(\bar{\phi}\big)}\Big(G\big(\bar{\phi}\big)_{,\eta}\Big)^2
+\frac{1}{2}\frac{1}{G\big(\bar{\phi}\big)}\Big(G\big(\bar{\phi}\big)_{,\eta\eta}\Big)
+\frac{1}{G\big(\bar{\phi}\big)}
\Bigg[\frac{G'\big(\bar{\phi}\big)}{G\big(\bar{\phi}\big)}\Big(-2F'\big(\bar{\phi}\big)V\big(\bar{\phi}\big)
-\frac{1}{2}F\big(\bar{\phi}\big)V'\big(\bar{\phi}\big)\Big)\nn\\
&&
+16\frac{\Big(F'\big(\bar{\phi}\big)\Big)^2}{F\big(\bar{\phi}\big)}V\big(\bar{\phi}\big)
+9F'\big(\bar{\phi}\big)V'\big(\bar{\phi}\big)
+4F''\big(\bar{\phi}\big)V\big(\bar{\phi}\big)
+F\big(\bar{\phi}\big)V''\big(\bar{\phi}\big)\Bigg].
\ea

 Introducing the auxiliary gauge invariant variables
 \ba
 &&v_{S}\equiv \frac{a\sqrt{G\big(\bar{\phi}\big)}}{F^{\frac{3}{2}}\big(\bar{\phi}\big)}
 \bigg(\delta \phi^{GI}+\frac{F^{2}\big(\bar{\phi}\big)}{\mathcal{G}}\bar{\phi}_{,\eta}\Psi\bigg),\quad
z_{S}\equiv \frac{a\sqrt{F\big(\bar{\phi}\big)G\big(\bar{\phi}\big)}}{\mathcal{G}}\bar{\phi}_{,\eta},\label{defvszs}
 \ea
 and using Eqs. (\ref{pdcgi})~(\ref{d2deltaphi}), after tedious calculation, we obtain
 \ba
v_{S,\eta\eta}
 -\frac{1}{F^4\big(\bar{\phi}\big)}\nabla^2v_{S}
 +\frac{2}{F\big(\bar{\phi}\big)}F\big(\bar{\phi}\big)_{,\eta}v_{S,\eta}
 -\frac{1}{z_{S}}\bigg[z_{S,\eta\eta}
 +\frac{2}{F\big(\bar{\phi}\big)}F\big(\bar{\phi}\big)_{,\eta}z_{S,\eta}\bigg]v_{S}=0.\label{SMuk}
 \ea

 It is not difficult to show that the Hamiltonian of gauge invariant scalar perturbations which can yield Eq. (\ref{SMuk}) is given by
 \ba
 \textbf{H}^{(2)}_S[\bar{N}]=\int_{\Sigma}d^3x \frac{\bar{N}}{2aF^2\big(\bar{\phi}\big)}\Bigg[\pi_S^2+\delta^{ab}(\partial_av_S)\partial_bv_S
 -\frac{F^4\big(\bar{\phi}\big)}{z_{S}}\bigg(z_{S,\eta\eta}
 +\frac{2}{F\big(\bar{\phi}\big)}F\big(\bar{\phi}\big)_{,\eta}(z_{S})_{,\eta}\bigg)v^2_{S}\Bigg],\label{HvS}
 \ea
 in which $\pi_S$ denotes the conjugate momentum of $v_S$ which satisfies $\{v_S(\vec{x}),\pi_S(\vec{y})\}=\delta^{(3)}(\vec{x}-\vec{y})$.

For tensor perturbation, we have $\delta q_{ab}=a^2h_{ab}$ where $h_{ab}$ is a symmetric trace-free and transversal tensor satisfying $\delta^{ab}\partial_{a}h_{bc}=\delta^{ab}h_{ab}=0$. Since from Eq. (\ref{gtq})  it is easy to see that the tensor perturbation is gauge invariant, we can define the auxiliary gauge invariant variables
\ba
 &&z_{T}\equiv \frac{a}{\sqrt{2\kappa F\big(\bar{\phi}\big)}},\quad
 v_{T}\equiv z_{T}h_{ab}.\label{defvtzt}
 \ea
 From the Hamilton's equations of perturbed variables, we obtain the equation of motion of $v_T$,
 \ba
 v_{T,\eta\eta}
 -\frac{1}{F^4\big(\bar{\phi}\big)}\nabla^2v_{T}
 +\frac{2}{F\big(\bar{\phi}\big)}F\big(\bar{\phi}\big)_{,\eta}v_{T,\eta}
 -\frac{1}{z_{T}}\bigg[z_{T,\eta\eta}
 +\frac{2}{F\big(\bar{\phi}\big)}F\big(\bar{\phi}\big)_{,\eta}z_{T,\eta}\bigg]v_{T}=0,\label{TMuk}
 \ea
which takes exactly the same form as Eq. (\ref{SMuk}) except that the subscript ``$S$" is replaced by ``$T$".  It can be shown that the Hamiltonian of tensor perturbations which can yield Eq. (\ref{TMuk}) is given by
\ba
\textbf{H}^{(2)}_T[\bar{N}]=\frac{1}{4}\int_{\Sigma}d^3x \frac{\bar{N}}{2aF^2\big(\bar{\phi}\big)}\Bigg[\pi_T^2+\delta^{ab}(\partial_av_T)\partial_bv_T
 -\frac{F^4\big(\bar{\phi}\big)}{z_{T}}\bigg(z_{T,\eta\eta}
 +\frac{2}{F\big(\bar{\phi}\big)}F\big(\bar{\phi}\big)_{,\eta}z_{T,\eta}\bigg)v^2_{T}\Bigg],\label{HvT}
\ea
in which $\pi_T$ denotes the conjugate momentum of $v_T$.

Note that the equations of motion of $v_{S}$ and $v_{T}$ can reproduce the Mukhonov equations of scalar and tensor perturbations of the minimally coupled case if $F\big(\bar{\phi}\big)=G\big(\bar{\phi}\big)=1$. For brevity, we also call Eqs. (\ref{SMuk}) and (\ref{TMuk}) the Mukhonov equations in the following sections.
\subsection{Causality}
As illustrated by Eqs. (\ref{SMuk}) and (\ref{TMuk}), the square of the propagation speed of perturbations satisfies
\ba
c^2_{S}=c^2_{T}=\frac{1}{F^4\big(\bar{\phi}\big)},\label{cssq}
\ea
where $c_{S}$ and $c_{T}$  denote the propagation speed of scalar perturbation and tensor perturbation respectively.
Thus,  we get $c_{S}>1$ when $F\big(\bar{\phi}\big)<1$ such that the speed of perturbations seems to become superluminal. However, considering that not only the propagation speed of the scalar and tensor perturbations can receive quantum gravity corrections but the propagation speed of electromagnetic fields can also be affected by quantum gravity effects, we should compare the propagation speed of the scalar and tensor perturbations with the physical speed (instead of the classical speed) of electromagnetic fields on the same quantum effective spacetime background.

In the case we consider, the Hamiltonian constraint of electromagnetic fields is given by \cite{Bojowald:2007a}
\ba
\textbf{C}_{b_{-}}^{(EM)}[N]=\int_{\Sigma}d^3x N\bigg[\alpha(\phi)\frac{2\pi}{\sqrt{\det(q)}}q_{ab}\pi^a\pi^b
+\beta(\phi)\frac{\sqrt{\det(q)}}{16\pi}q^{ac}q^{bd}F_{ab}F_{bd}\bigg],\label{qcHcem}
\ea
in which the functions $\alpha(\phi)$ and $\beta(\phi)$ denote the undetermined quantum corrections. In the classical case, we have $\alpha=\beta=1$. The conjugate variables in (\ref{qcHcem}) are  the spatial component of the vector potential $A_a$ and its conjugate momentum $\pi^a$. The spatial component of the field strength tensor is defined by $F_{ab}\equiv\partial_aA_b-\partial_bA_a$.

The diffeomorphism constraint of electromagnetic fields is given by \cite{Bojowald:2007a}
\ba
\textbf D^{(EM)}[N^a]=\int_{\Sigma}d^3x N^a\pi^bF_{ab}.\label{qcDcem}
\ea
Note that we require that like the gravitational part the diffeomorphism constraint of electromagnetic fields does not receive quantum corrections either.
Now, the total Hamiltonian constraint and diffeomorphism constraint read, respectively, as
\ba
\textbf{C}_{b_{-}}^{(total)}[N]&=&\textbf{C}_{b_{-}}[N]+\textbf{C}_{b_{-}}^{(EM)}[N],\label{Ctotal}\\
\textbf{D}^{(total)}[N^a]&=&\textbf{D}[N^a]+\textbf{D}^{(EM)}[N^a].
\ea
Straightforward calculation gives
\ba
\big\{\textbf{C}^{(total)}_{b_{-}}[M],\textbf{C}^{(total)}_{b_{-}}[N]\big\}
=\textbf{D}\bigg[\frac{1}{F^4(\phi)}q^{ab}(MD_bN-ND_bM)\bigg]
+\textbf{D}^{(EM)}\bigg[\alpha(\phi)\beta(\phi)q^{ab}(MD_bN-ND_bM)\bigg].
\ea
To obtain a first class constraint algebra, we should require
\ba
\alpha(\phi)\beta(\phi)=\frac{1}{F^4(\phi)}.\label{alphabeta}
\ea
As shown in Ref. \cite{Bojowald:2007b}, the group velocity of electromagnetic wave propagating on the spatially flat FRW background is
\ba
c_{EM}=\sqrt{\bar{\alpha}\bar{\beta}},\label{gvEM}
\ea
where $\bar{\alpha}$ and $\bar{\beta}$ denote the background value of $\alpha(\phi)$ and $\beta(\phi)$ respectively.  Then, using Eqs. (\ref{alphabeta}) and (\ref{cssq}),  we obtain
\ba
c_{EM}=c_{S}=c_{T}=\frac{1}{F^2\big(\bar{\phi}\big)},\label{cemeqcs}
\ea
which indicates that the causality is still respected by the quantum corrections.
\section{Solutions of the Mukhanov equations under slow-roll approximation}
In this section, we  solve the Mukhanov equations under slow-roll approximation to obtain spectral indices of the perturbations. To justify this practice, we assume that the slow-roll inflation can take place, and for any wave number in the present observational range the energy density at the instant of the horizon crossing is significantly lower than the critical energy density of LQC .

Due to their complex forms, it is difficult to directly solve the Mukhanov equations (\ref{SMuk}) and (\ref{TMuk}). For convenience, we define a new time variable $d\zeta$ which relates to the conformal time by
\ba
d\zeta=\frac{1}{F^2(\bar{\phi})}d\eta.\label{rct}
\ea
Since the conformal time $d\eta$ corresponds to the choice of the lapse function $\bar{N}=a$, it is easy to see that the new time variable $d\zeta$ corresponds to the choice of the lapse function $\bar{N}=aF^2\big(\bar{\phi}\big)$.
Using this new variable, the Mukhanov equations of scalar perturbation and tensor perturbation can be reexpressed in a relatively simple form,
\ba
 \frac{d^2v_{S,T}}{d\zeta^2}
 -\nabla^2v_{S,T}
 -\bigg(\frac{1}{z_{S,T}}\frac{d^2z_{S,T}}{d\zeta^2}\bigg) v_{S,T}=0;\label{MukEST}
\ea
and the perturbed Hamiltonian (\ref{HvS}) and (\ref{HvT}) can also be rewritten in a simple form,
\ba
 \textbf{H}^{(2)}_{S,T}=\Upsilon_{S,T}\int_{\Sigma}d^3x \bigg[\pi_{S,T}^2+\delta^{ab}(\partial_av_{S,T})\partial_bv_{S,T}
 -\bigg(\frac{1}{z_{S,T}}\frac{d^2z_{S,T}}{d\zeta^2}\bigg)v^2_{S,T}\bigg],\label{HMukST}
\ea
where $\Upsilon_{S}=1$ and $\Upsilon_{T}=\frac{1}{4}$.

  During the slow-roll period, both the scalar field and the Hubble parameter vary very slowly with respect to the proper time. It is useful to introduce the four slow-roll parameters,
 \ba
 \epsilon_1=\frac{\dot{H}}{H^2},\quad\epsilon_2=\frac{\ddot{\bar{\phi}}}{H\dot{\bar{\phi}}},
 \quad\epsilon_3=\frac{\dot{F}\big(\bar{\phi}\big)}{2HF\big(\bar{\phi}\big)},
 \quad\epsilon_4=\frac{\dot{G}\big(\bar{\phi}\big)}{2HG\big(\bar{\phi}\big)},\label{srp}
 \ea
and the slow-roll condition is satisfied if $\epsilon_i\ll1$ for all $\epsilon_i$.

 Using the relation $dt=ad\eta=aF^2\big(\bar{\phi}\big)d\zeta$ and the definition of $z_S$  in Eq. (\ref{defvszs}), we have
 \ba
 v_{S}= \frac{a\sqrt{G\big(\bar{\phi}\big)}}{F^{\frac{3}{2}}\big(\bar{\phi}\big)}
 \Bigg(\delta \phi^{GI}
 +\frac{\dot{\bar{\phi}}}{\Big(H-\frac{\dot{F}(\bar{\phi})}{2F(\bar{\phi})}\Big)}\Psi\Bigg),\quad
 z_S=
 \frac{a\sqrt{G\big(\bar{\phi}\big)}\dot{\bar{\phi}}}
 {F^{\frac{3}{2}}\big(\bar{\phi}\big)\bigg(H-\frac{\dot{F}(\bar{\phi})}{2F(\bar{\phi})}\bigg)};\label{zsdt}
 \ea
 then, using the definition of slow-roll parameters we obtain
 \ba
 \frac{1}{z_{S}}\frac{dz_{S}}{d\zeta}&\simeq& F^2\big(\bar{\phi}\big)aH(1-\epsilon_1+\epsilon_2-3\epsilon_3+\epsilon_4),\label{appdzs}\\
 \frac{1}{z_{S}}\frac{d^2z_{S}}{d\zeta^2}&\simeq& F^4\big(\bar{\phi}\big)a^2H^2(2-2\epsilon_1+3\epsilon_2-5\epsilon_3+3\epsilon_4).\label{appddzs}
 \ea
 For simplicity, in Eqs. (\ref{appdzs}) and (\ref{appddzs}) both the time variation and higher order terms of  $\epsilon_i$ have been neglected; and the same is done in the following calculations.

 From the equation
 \ba
 \frac{d}{d\zeta}\bigg(\frac{1}{F^2(\bar{\phi})aH}\bigg)\simeq-(1+\epsilon_1+4\epsilon_3),\label{dteta}
 \ea
 we obtain
 \ba
 F^4\big(\bar{\phi}\big)a^2H^2\simeq\frac{1}{\zeta^2}\frac{1}{1+2\epsilon_1+8\epsilon_3}.\label{appteta}
 \ea
 Substituting Eq. (\ref{appteta}) into Eq. (\ref{appddzs}), we get
 \ba
 \frac{1}{z_{S}}\frac{d^2z_{S}}{d\zeta^2}\simeq\frac{m_S}{\zeta^2}, \quad m_S\equiv2-6\epsilon_1+3\epsilon_2-21\epsilon_3+3\epsilon_4.\label{ms}
 \ea

 Similarly, for tensor perturbation, we obtain
 \ba
 \frac{1}{z_{T}}\frac{d^2z_{T}}{d\zeta^2}\simeq \frac{m_T}{\zeta^2},\quad
 m_T\equiv2-3\epsilon_1-15\epsilon_3.\label{mt}
 \ea
 Hence, in the slow-roll period, the Mukhanov equations can be approximately written as
 \ba
 \frac{d^2v_{S,T}}{d\zeta^2}
 -\nabla^2v_{S,T}
 -\frac{(\nu^2_{S,T}-1/4)}{\zeta^2}v_{S,T}=0,\label{MukEST}
 \ea
 where in order to proceed we have introduced the variables $\nu_{S}$ and $\nu_{T}$,
 \ba
 \nu_{S}\equiv\sqrt{m_{S}+\frac{1}{4}}&\simeq&\frac{3}{2}-2\epsilon_1+\epsilon_2-7\epsilon_3+\epsilon_4,\label{nuS}\\
 \nu_{T}\equiv\sqrt{m_{S}+\frac{1}{4}}&\simeq&\frac{3}{2}-\epsilon_1-5\epsilon_3.\label{nuT}
 \ea
\subsection{Spectral indices of the slow-roll inflation}
 Now, let us solve the Mukhanov equation of scalar perturbation first. The treatment mimics that in classical theory.
First, we promote the quantities $v_S$ and $\pi_S$ to quantum operators which satisfy the equal time commutation relation,
\ba
\big[\hat{v}_S(\zeta,\vec{x}),\hat{\pi}_S(\zeta,\vec{y})\big]=i\delta^{(3)}\big(\vec{x}-\vec{y}\big),
\quad \big[\hat{v}_S(\zeta,\vec{x}),\hat{v}_S(\zeta,\vec{y})\big]
=\big[\hat{\pi}_S(\zeta,\vec{x}),\hat{\pi}_S(\zeta,\vec{y})\big]=0,\label{crpv}
\ea
where we have set $\hbar=1$, and using the Heisenberg's equation of motion, we get
\ba
\frac{d\hat{v}_S}{d\zeta}=\frac{1}{i}\Big[\hat{v}_S,\textbf{H}^{(2)}_{S}\Big]=\hat{\pi}_S.\label{dvs}
\ea
Then, we Fourier decompose $\hat{v}_S$ as
\ba
\hat{v}_S(\zeta,\vec{x})
=\frac{1}{(2\pi)^{\frac{3}{2}}}\int d^3\vec{k}\bigg[\hat{a}_S\big(\vec{k}\big)v_{Sk}(\zeta)e^{i\vec{k}\cdot\vec{x}}
+\hat{a}^{\dagger}_S\big(\vec{k}\big)v^{*}_{Sk}(\zeta)e^{-i\vec{k}\cdot\vec{x}}\bigg],\label{Fdvs}
\ea
where $k\equiv|\vec{k}|$.
Plugging Eq. (\ref{Fdvs}) into Eqs. (\ref{crpv}) and (\ref{dvs}) and requiring $\big[\hat{a}_S\big(\vec{k}_1\big), \hat{a}^{\dagger}_S\big(\vec{k}_2\big)\big]=\delta^{(3)}\big(\vec{k}_1-\vec{k}_2\big)$,  we obtain
\ba
v_{\textit{Sk}}\frac{dv^{*}_{Sk}}{d\zeta}-v^{*}_{Sk}\frac{dv_{Sk}}{d\zeta}=i.\label{wrc}
\ea
  With the help of Eqs. (\ref{MukEST}) and (\ref{Fdvs}), we obtain the evolution equation of $v_{Sk}$,
 \ba
 \frac{d^2v_{Sk}}{d\zeta^2}+\Bigg[k^2-\frac{(\nu^2_{S}-1/4)}{\zeta^2}\Bigg]v_{Sk}=0.\label{SMukk}
 \ea
The solution for $v_{Sk}$ is given by
\ba
v_{Sk}(\zeta)=\frac{\sqrt{\pi|\zeta|}}{2}
\big[c_1\big(\vec{k}\big)H^{(1)}_{\nu_S}\big(k|\zeta|\big)
+c_2\big(\vec{k}\big)H^{(2)}_{\nu_S}\big(k|\zeta|\big)\big],\label{svsk}
\ea
where $H_{\nu}^{(1,2)}$ are Hankel functions. To determine the coefficients $c_1\big(\vec{k}\big)$ and $c_2\big(\vec{k}\big)$, we can use the asymptotic property of the Hankel functions in the limit of small scales where $k|\zeta|\gg1$,
\ba
v_{Sk}(\zeta)\big|_{k|\zeta|\gg1}\simeq
\frac{1}{\sqrt{2k}}\Big[c_1\big(\vec{k}\big)e^{-i\frac{\pi}{4}(1+2\nu_S)}e^{ik|\zeta|}
+c_2\big(\vec{k}\big)e^{i\frac{\pi}{4}(1+2\nu_S)}e^{-ik|\zeta|}\Big].\label{vskssl}
\ea
Substituting Eq. (\ref{vskssl}) into Eq. (\ref{wrc}), we obtain
\ba
\big|c_1\big(\vec{k}\big)\big|^2-\big|c_2\big(\vec{k}\big)\big|^2=1.\label{wrcss}
\ea
Assuming that only the positive frequency solution remains in the small scale limit for the wave numbers which lie in the current observational range, we can set
\ba
c_1\big(\vec{k}\big)=1,\quad c_2\big(\vec{k}\big)=0,\label{c1c2}
\ea
which corresponds to selecting the Bunch-Davies vacuum for these wave numbers when $|\zeta|\gg\frac{1}{k}$.

Using the asymptotic property of the Hankel functions in the large scale limit $k|\zeta|\ll1$, we obtain
\ba
v_{Sk}(\zeta)\big|_{k|\zeta|\ll1}\simeq
\frac{-i}{2\sqrt{\pi}}\bigg(\frac{2}{k}\bigg)^{\nu_S}\Gamma(\nu_S)|\zeta|^{\frac{1}{2}-\nu_S}.\label{vsklsl}
\ea

Introducing the variable $\mathcal{R}_k(\zeta)\equiv\frac{v_{Sk}}{z_S}$, from the definition of the power spectrum of scalar perturbation $P_{\mathcal{R}}(k,\zeta)\equiv\frac{k^3}{2\pi^2}|\mathcal{R}_k(\zeta)|^2$,  we get
\ba
P_{\mathcal{R}}(k,\zeta)= \frac{2^{2\nu_S}\Gamma^2(\nu_S)}{\pi^3}\frac{|\zeta|^{1-2\nu_S}}{z^2_S}
\bigg(\frac{k}{2}\bigg)^{3-2\nu_S}.\label{PR}
\ea
 Moreover, from the equation
\ba
\frac{1}{z_{S}}\frac{d^2z_{S}}{d\zeta^2}\simeq\frac{\nu^2_S-\frac{1}{4}}{\zeta^2},\label{sdzs}
\ea
we obtain
\ba
z_S\propto|\zeta|^{\frac{1}{2}-\nu_S};\label{zseta}
\ea
 plugging Eq. (\ref{zseta}) into Eq. (\ref{PR}), we find that $P_{\mathcal{R}}$ becomes time independent in the large scale limit.

 From Eq. (\ref{zseta}), we get
\ba
z_S=z_S^*\bigg|\frac{\zeta}{\zeta^*}\bigg|^{\frac{1}{2}-\nu_S},\label{zsstar}
\ea
in which $\zeta^*$ denotes an arbitrary instant and $z_S^*\equiv z_S\big|_{\zeta=\zeta^*}$. Substituting Eq. (\ref{zsstar}) into Eq. (\ref{PR}), we have
\ba
P_{\mathcal{R}}\simeq\frac{1}{4\pi^2}\frac{|k\zeta^*|^{3-2\nu_S}}{(z_S^*)^2(\zeta^*)^2},\label{PRapp}
\ea
where we have used $\Gamma(\nu_S)\simeq\Gamma(\frac{3}{2})$. The above expression can be simplified by choosing the instant $|\zeta^*|=\frac{1}{k}$. Note that at this instant we also have $F^2\big(\bar{\phi}\big)aH=k$ because from  Eq. (\ref{appteta}) we learn that $\frac{1}{|\zeta|}\simeq F^2\big(\bar{\phi}\big)aH$. Then, substituting Eq. (\ref{zsdt}) into Eq. (\ref{PRapp}), we find that $P_{\mathcal{R}}$ can also be expressed as
\ba
P_{\mathcal{R}}\simeq
\frac{1}{4\pi^2}\frac{H^4F^7(\bar{\phi})}{G(\bar{\phi})(\dot{\bar{\phi}})^2}\Bigg|_{k=F^2(\bar{\phi})aH}.
\ea
Now, the spectral index of scalar perturbation is given by
\ba
n_S-1\equiv\frac{d\ln P_{\mathcal{R}}}{d\ln k}\bigg|_{k=F^2(\bar{\phi})aH}=
\bigg(\frac{1}{d\ln k/d t}
\frac{d\ln P_{\mathcal{R}}}{dt}\bigg)\bigg|_{k=F^2(\bar{\phi})aH}
\simeq(4\epsilon_1-2\epsilon_2+14\epsilon_3-2\epsilon_4)\big|_{k=F^2(\bar{\phi})aH}.\label{nS}
\ea

For tensor perturbation, we find that the solution for $v_{Tk}(\zeta)$ takes the same form as $v_{Sk}(\zeta)$ except that $\nu_S$ should be replaced by $\nu_T$. Introducing the variable $h_k\equiv\frac{v_{Tk}}{z_T}$ and using the definition of the power spectrum of tensor perturbation $P_{h}(k,\zeta)\equiv\frac{2k^3}{\pi^2}|h_k|^2$, then simply following the above treatment for  scalar perturbation, we obtain
\ba
P_{h}\equiv\frac{2k^3}{\pi^2}|h_k|^2\simeq \frac{2\kappa}{\pi^2}H^2F^5(\bar{\phi})\Bigg|_{k=F^2(\bar{\phi})aH},
\ea
from which the  spectral index of tensor perturbation can be read as,
\ba
n_T\equiv\frac{d\ln P_{h}}{d\ln k}\simeq(2\epsilon_1+10\epsilon_3)\big|_{k=F^2(\bar{\phi})aH},\label{nT}\label{nT}
\ea
where we have used Eq. (\ref{nuT}) in the last step. Moreover, it is easy to show that $z_T\propto|\zeta|^{\frac{1}{2}-\nu_T}$; thus, $P_h$ also becomes time independent in the large scale limit.

The tensor-to-scalar ratio now reads
\ba
r\equiv \frac{P_{h}}{P_{\mathcal{R}}}
\simeq8\kappa\frac{G(\bar{\phi})(\dot{\bar{\phi}})^2}{H^2F^2}\Bigg|_{k=F^2(\bar{\phi})aH}
\simeq-16(\epsilon_1+5\epsilon_3)\big|_{k=F^2(\bar{\phi})aH}.\label{tts}
\ea
where  in the last step we have used the background equation of motion (\ref{qRaeq}) in the limit $\cos b\rightarrow-1$ along with the slow-roll condition. Comparing Eq. (\ref{tts}) with Eq. (\ref{nT}), we find $r\simeq-8n_T$.

It is worth mentioning that in the Jordan frame of classical STT the spectral indices are given by \cite{Faraoni:2004}
\ba
n_S-1\simeq(4\epsilon_1-2\epsilon_2+2\epsilon_3-2\epsilon_4)\big|_{k=aH},\quad
n_T\simeq(2\epsilon_1-2\epsilon_3)\big|_{k=aH},\quad
r\simeq-8n_T.\label{spicstt}
\ea
Obviously, the spectral indices in the classical case differ from the spectral indices in the case $\cos b\rightarrow-1$ in two aspects: the coefficient before $\epsilon_3$ and the instant at which they take value. Nevertheless, the consistency relation between the tensor-to-scalar ratio and the tensor spectral index remains the same in both cases.

Under the slow-roll approximation, the background equations in the Jordan frame can be approximated as
\ba
H^2\simeq \frac{\kappa}{3}\frac{V(\bar{\phi})}{F(\bar{\phi})},\quad
3H\dot{\bar{\phi}}\simeq-\frac{4F'\big(\bar{\phi}\big)V\big(\bar{\phi}\big)+F\big(\bar{\phi}\big)V'\big(\bar{\phi}\big)}{G\big(\bar{\phi}\big)}.\label{FrV}
\ea

Using (\ref{FrV}), the power spectrum of the scalar perturbations can be expressed in terms of $\bar{\phi}$,
\ba
P_{\mathcal{R}}\simeq
\frac{1}{4\pi^2}\frac{H^4F^7(\bar{\phi})}{G(\bar{\phi})(\dot{\bar{\phi}})^2}\Bigg|_{k=F^2(\bar{\phi})aH}
\simeq\frac{\kappa^3}{12\pi^2}
\frac{F^{10}(\bar{\phi})G(\bar{\phi})V^3(\bar{\phi})}{\big(4F^3(\bar{\phi})F'(\bar{\phi})V(\bar{\phi})+F^4(\bar{\phi})V(\bar{\phi})\big)^2}\Bigg|_{k=F^2(\bar{\phi})aH}.
\label{PRs}
\ea

Direct calculations show that the slow-roll parameters can be expressed as
\ba
&&\epsilon_1\simeq2\epsilon_{F}-\frac{3}{2}\epsilon_{FV}-\frac{1}{2}\epsilon_{V},\quad
\epsilon_2\simeq-2\epsilon_{F}-\frac{7}{2}\epsilon_{FV}+\frac{1}{2}\epsilon_{V}+4\epsilon_{FG}
+\epsilon_{GV}-4\eta_{F}-\eta_{V},\nn\\
&&\epsilon_3\simeq-2\epsilon_{F}-\frac{1}{2}\epsilon_{FV},\quad
\epsilon_4\simeq-2\epsilon_{FG}-\frac{1}{2}\epsilon_{GV},\label{epsilons}
\ea
where
\ba
&&\epsilon_{F}\equiv\frac{1}{\kappa}\frac{\big(F'(\bar{\phi})\big)^2}{G(\bar{\phi})},\quad
\epsilon_{FV}\equiv\frac{1}{\kappa}\frac{F(\bar{\phi})F'(\bar{\phi})V'(\bar{\phi})}{G(\bar{\phi})V(\bar{\phi})},\quad
\epsilon_{V}\equiv\frac{1}{\kappa}\frac{F^2(\bar{\phi})\big(V'(\bar{\phi})\big)^2}{G(\bar{\phi})V^2(\bar{\phi})},\quad
\epsilon_{FG}\equiv\frac{1}{\kappa}\frac{F(\bar{\phi})F'(\bar{\phi})G'(\bar{\phi})}{G^2(\bar{\phi})},\nn\\
&&\epsilon_{GV}\equiv\frac{1}{\kappa}\frac{F^2(\bar{\phi})G'(\bar{\phi})V'(\bar{\phi})}{G^2(\bar{\phi})V(\bar{\phi})},\quad
\eta_{F}\equiv\frac{1}{\kappa}\frac{F(\bar{\phi})F''(\bar{\phi})}{G(\bar{\phi})},\quad
\eta_{V}\equiv\frac{1}{\kappa}\frac{F^2(\bar{\phi})V''(\bar{\phi})}{G(\bar{\phi})V(\bar{\phi})},
\ea
such that the indices in the Jordan frame can also be given by
\ba
n_S-1&\simeq&
-(16\epsilon_{F}+6\epsilon_{FV}+3\epsilon_{V}+4\epsilon_{FG}+\epsilon_{GV}-8\eta_{F}-2\eta_{V})\big|_{k=F^2(\bar{\phi})aH},\label{nSs}\\
r&\simeq&16(8\epsilon_{F}+4\epsilon_{FV}+\epsilon_{V})\big|_{k=F^2(\bar{\phi})aH}.\label{rs}
\ea

\subsection{Comparison with the results in the Einstein frame}
The analysis in the previous subsection is performed in the Jordan frame. In this subsection, we check whether the results are equivalent to those in the Einstein frame.

In the following, we drop a ``tilde" to denote the variables in the Einstein frame. From the transformation of the variables in (\ref{canotrans}), we find that the background and perturbed variables in the Einstein frame are related to their counterparts in the Jordan frame by
\ba
\tilde{a}=\frac{a}{\sqrt{F\big(\bar{\phi}\big)}}, \quad d\tilde{t}=\frac{1}{F^{\frac{5}{2}}\big(\bar{\phi}\big)}dt,\quad
\tilde{H}\equiv\frac{1}{\tilde{a}}\frac{d \tilde{a}}{d \tilde{t}}=F^{\frac{5}{2}}\big(\bar{\phi}\big)
\Bigg(H-\frac{\dot{F}\big(\bar{\phi}\big)}{2F\big(\bar{\phi}\big)}\Bigg),\label{rebEJ}
\ea
and
\ba
\delta \tilde{\phi}= \frac{\sqrt{G(\phi)}}{F(\phi)}\delta \phi,\quad
\tilde{\varphi}=\varphi-\frac{5}{2}\frac{F'\big(\bar{\phi}\big)}{F\big(\bar{\phi}\big)}\delta \phi,\quad
\tilde{\psi}=\psi-\frac{1}{2}\frac{F'\big(\bar{\phi}\big)}{F\big(\bar{\phi}\big)}\delta \phi,\quad
\tilde{E}=E,\quad
\tilde{h}_{ab}=h_{ab}.\label{reperEJ}
\ea

From Eqs. (\ref{defPsi}), (\ref{defdeltaphiGI}), (\ref{defmathG}) and (\ref{defvszs}), we find that the curvature perturbation $\mathcal{R}$ in the Jordan frame satisfies
\ba
\mathcal{R}\equiv\frac{v_S}{z_S}=
\psi-\frac{1}{2}\frac{F'\big(\bar{\phi}\big)}{F\big(\bar{\phi}\big)}\delta \phi
+\frac{H-\frac{\dot{F}(\bar{\phi})}{2F(\bar{\phi})}}{\dot{\bar{\phi}}}\delta \phi.
\ea
Recall that in the Einstein frame the curvature perturbation is defined by
\ba
\tilde{\mathcal{R}}\equiv\tilde{\psi}+\frac{\tilde{H}}{d\bar{\tilde{\phi}}/d\tilde{t}}\delta \tilde{\phi}.
\ea
With the help of Eqs. (\ref{rebEJ}) and (\ref{reperEJ}), it is easy to find $\tilde{\mathcal{R}}=\mathcal{R}$. Furthermore,
it can be shown that the power spectrum $\tilde{P}_{\tilde{\mathcal{R}}}$ in the Einstein frame satisfies
 \ba
 \tilde{P}_{\tilde{\mathcal{R}}}
 \simeq\frac{1}{4\pi^2}\frac{\tilde{H}^4}{(d\bar{\tilde{\phi}}/d\tilde{t})^2}\Bigg|_{k=\tilde{a}\tilde{H}}
 =\frac{1}{4\pi^2}\frac{F^7(\bar{\phi})H^4(1-\epsilon_3)^4}{G(\bar{\phi})(\dot{\bar{\phi}})^2}
 \Bigg|_{k=F^2(\bar{\phi})\tilde{a}\tilde{H}(1-\epsilon_3)}\simeq P_{\mathcal{R}}.
 \ea

 Moreover, since the tensor perturbation is invariant under the conformal transformation, we have $\tilde{h}_{k}=h_{k}$, and the power spectrum $\tilde{P}_{\tilde{h}}$ in the Einstein frame satisfies
\ba
\tilde{P}_{\tilde{h}}\simeq\frac{2\kappa}{\pi^2}\tilde{H}^2\bigg|_{k=\tilde{a}\tilde{H}}
=\frac{2\kappa}{\pi^2}F^5(\bar{\phi})H^2(1-\epsilon_3)^2\bigg|_{k=F^2(\bar{\phi})aH(1-\epsilon_3)}
\simeq P_{h}.
\ea
In the Einstein frame, to linear order of the slow-roll parameters, the spectral index of the scalar perturbation is expressed by
\ba
\tilde{n}_S-1\equiv\frac{d\ln \tilde{P}_{\tilde{\mathcal{R}}}}{d\ln k}\simeq4\tilde{\epsilon}_1-2\tilde{\epsilon}_2,\label{tildens}
\quad\tilde{n}_T\equiv\frac{d\ln \tilde{P}_{\tilde{h}}}{d\ln k}\simeq2\tilde{\epsilon}_1,
\ea
and using Eq. (\ref{rebEJ}), we find that
\ba
\tilde{\epsilon}_1&\equiv&\frac{1}{\tilde{H}^2}\frac{d\tilde{H}}{d\tilde{t}}
=\frac{\frac{5}{2}F^4\big(\bar{\phi}\big)H\dot{F}\big(\bar{\phi}\big)
+F^5\big(\bar{\phi}\big)[\dot{H}(1-\epsilon_3)-H\dot{\epsilon}_3]}
{F^5\big(\bar{\phi}\big)H^2(1-\epsilon_3)^2}\bigg|_{k=\tilde{a}\tilde{H}}
\simeq(\epsilon_1+5\epsilon_3)\bigg|_{k=F^2(\bar{\phi})aH},\label{tildeep1}\\
\tilde{\epsilon}_2&\equiv&
\frac{1}{\tilde{H}}\frac{d^2\bar{\tilde{\phi}}/d\tilde{t}^2}{d\bar{\tilde{\phi}}/d\tilde{t}}
=\frac{\frac{3}{2}G\big(\bar{\phi}\big)\dot{\bar{\phi}}\dot{F}\big(\bar{\phi}\big)+\frac{1}{2}F\big(\bar{\phi}\big)\dot{G}\big(\bar{\phi}\big)\dot{\bar{\phi}}+F\big(\bar{\phi}\big)G\big(\bar{\phi}\big)\ddot{\bar{\phi}}}
{F\big(\bar{\phi}\big)G\big(\bar{\phi}\big)H(1-\epsilon_3)\dot{\bar{\phi}}}\bigg|_{k=\tilde{a}\tilde{H}}
\simeq(\epsilon_2+3\epsilon_3+\epsilon_4)\bigg|_{k=F^2(\bar{\phi})aH}.\label{tildeep2}
\ea
Substituting Eqs. (\ref{tildeep1}) and (\ref{tildeep2}) into Eq. (\ref{tildens}), we can easily show that $\tilde{n}_S\simeq n_S$ and $\tilde{n}_S\simeq n_T$. Moreover, the tensor-to-scalar ratio in the Einstein frame satisfies $\tilde{r}\simeq -8\tilde{n}_T\simeq r$. Thus, we conclude that the power spectra and spectral indices in the Einstein frame coincide with the ones in the Jordan frame to linear order of slow-roll parameters.
However, we should mention that further calculation shows that the coincidence of the results between the two frames does not hold to higher orders of slow-roll parameters.

\section{Cosmological dynamics of a specific model of STT}
In this section, we apply the results obtained in the previous sections to study a specific model of STT. To be specific, in action (\ref{STTaction}) we choose
\ba
F(\phi)=\frac{1}{(1+\xi\kappa\phi^2)^{\frac{1}{2}}},\quad K(\phi)=1,\quad V(\phi)=\frac{\lambda}{4}\phi^4,
\label{FKV}
\ea
in which the dimensionless coupling parameters $\xi$ and $\lambda$ are set to be greater than zero.

The background equations of motion are as follows:
\ba
&&\bigg(\frac{H}{(1+\xi\kappa\bar{\phi}^2)^{\frac{1}{2}}}
-\frac{1}{2}\frac{\xi\kappa\bar{\phi}\dot{\bar{\phi}}}{(1+\xi\kappa\bar{\phi}^2)^{\frac{3}{2}}}\cos b\bigg)^2
=\frac{\kappa}{3}\rho_e\bigg(1-\frac{\rho_e}{\rho_c}\bigg),\label{qFrxi}\\
&&\ddot{\bar{\phi}}+3H\dot{\bar{\phi}}
+\frac{1}{2}\frac{\dot{G}\big(\bar{\phi}\big)}{G\big(\bar{\phi}\big)}\dot{\bar{\phi}}
-\frac{-\frac{3}{4}\lambda\xi\kappa\bar{\phi}^5(\cos b+1)-\lambda\bar{\phi}^3}{(1+\xi\kappa\bar{\phi}^2)^{\frac{3}{2}}G\big(\bar{\phi}\big)}=0,\label{qKGxi}\\
&&\cos^2b=1-\frac{\rho_e}{\rho_c}, \label{cos2bxi}
\ea
where
\ba
\rho_e\equiv\frac{G\big(\bar{\phi}\big)}{2}\big(\dot{\bar{\phi}}\big)^2
+\frac{\lambda\bar{\phi}^4}{4(1+\xi\kappa\bar{\phi}^2)^{\frac{1}{2}}},\quad G\big(\bar{\phi}\big)\equiv\frac{3}{2}\frac{\xi^2\kappa\bar{\phi}^2}{(1+\xi\kappa\bar{\phi}^2)^3}
+\frac{1}{(1+\xi\kappa\bar{\phi}^2)^\frac{1}{2}}.\label{rhoexi}
\ea

Defining $\chi\equiv\dot{\bar{\phi}}$, the dynamical equations of the scalar field read
\ba
\frac{d\tilde{\phi}}{dt}&=&\chi,\label{dephi1}\\
\frac{d\chi}{dt}&=&-3H\chi-\frac{1}{2}\frac{\dot{G}\big(\bar{\phi}\big)}{G\big(\bar{\phi}\big)}\chi
-\frac{\frac{3}{4}\lambda\xi\kappa\bar{\phi}^5(\cos b+1)+\lambda\bar{\phi}^3}{(1+\xi\kappa\bar{\phi}^2)^{\frac{3}{2}}G\big(\bar{\phi}\big)},\label{dexi}
\ea
where $H$ and $\cos b$ are understood as functions of $\phi$ and $\chi$ via Eqs. (\ref{qFrxi}) and (\ref{cos2bxi}). Since $\xi>0$, $\lambda>0$, we have $F\big(\bar{\phi}\big)>0$, $K\big(\bar{\phi}\big)>0$ and $V\big(\bar{\phi}\big)\geq0$ for arbitrary values of $\bar{\phi}$. Moreover, from Eqs. (\ref{dephi1}) and (\ref{dexi}), we find that the dynamical system has only two fixed points: ($\bar{\phi}=0$,$\chi=0$) with $\cos b=1$ and ($\phi=0$,$\chi=0$) with $\cos b=-1$. According to the arguments in Sec. II, the phase space is naturally divided into two disconnected sectors by $\sin b<0$ and $\sin b>0$. In the former sector, the fixed point ($\bar{\phi}=0$,$\chi=0$) with $\cos b=1$ is the source of the system and the fixed point ($\bar{\phi}=0$,$\chi=0$) with $\cos b=-1$ is the sink of the system, while in the other sector the case is just the opposite. Furthermore, considering that the effective energy density is bounded from above by $\rho_c$, none of the phase space trajectories of solutions of equations can approach infinity in the phase space. Thus, if there are no limit circles in the sector $\sin b<0$, all trajectories of solutions starting from the the source in the asymptotic past will evolve to the sink  in the asymptotic future, which implies that each phase space trajectory passes through the bounce during the evolution. Hence, in this model, a classical contracting universe in the remote past can successfully evolve into an expanding universe described by the $b_{-}$ branch of equations via the bounce. In addition, since $F\big(\bar{\phi}\big)=1$ at the fixed point, we find that general relativity is the attractor of this model.

\begin{figure}[h]
\centering
\subfloat[\label{fig1a}
]{\includegraphics[height=5.5cm]{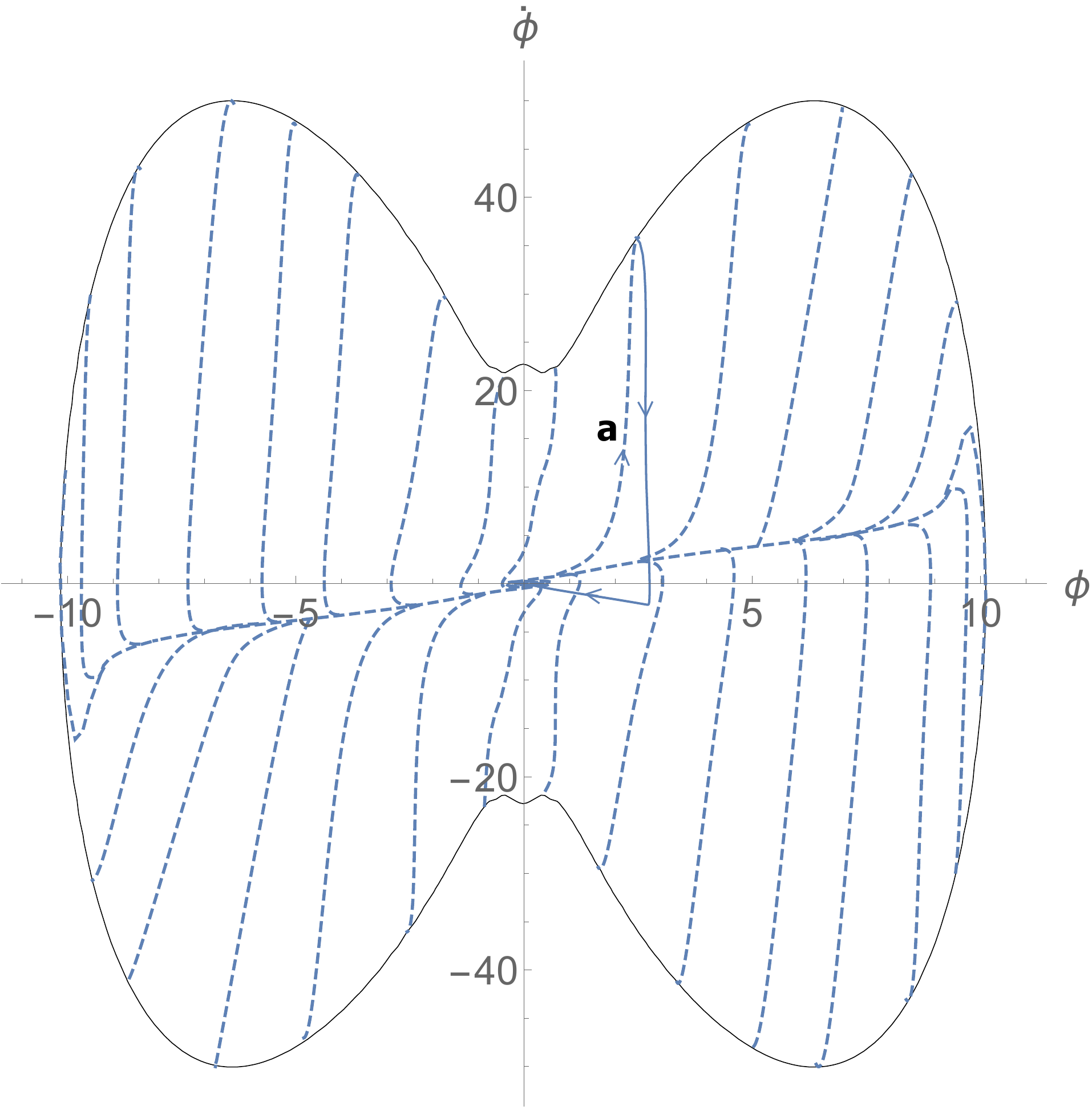}}
\hspace{1cm}
\subfloat[\label{fig1b}]{\includegraphics[height=5.2cm]{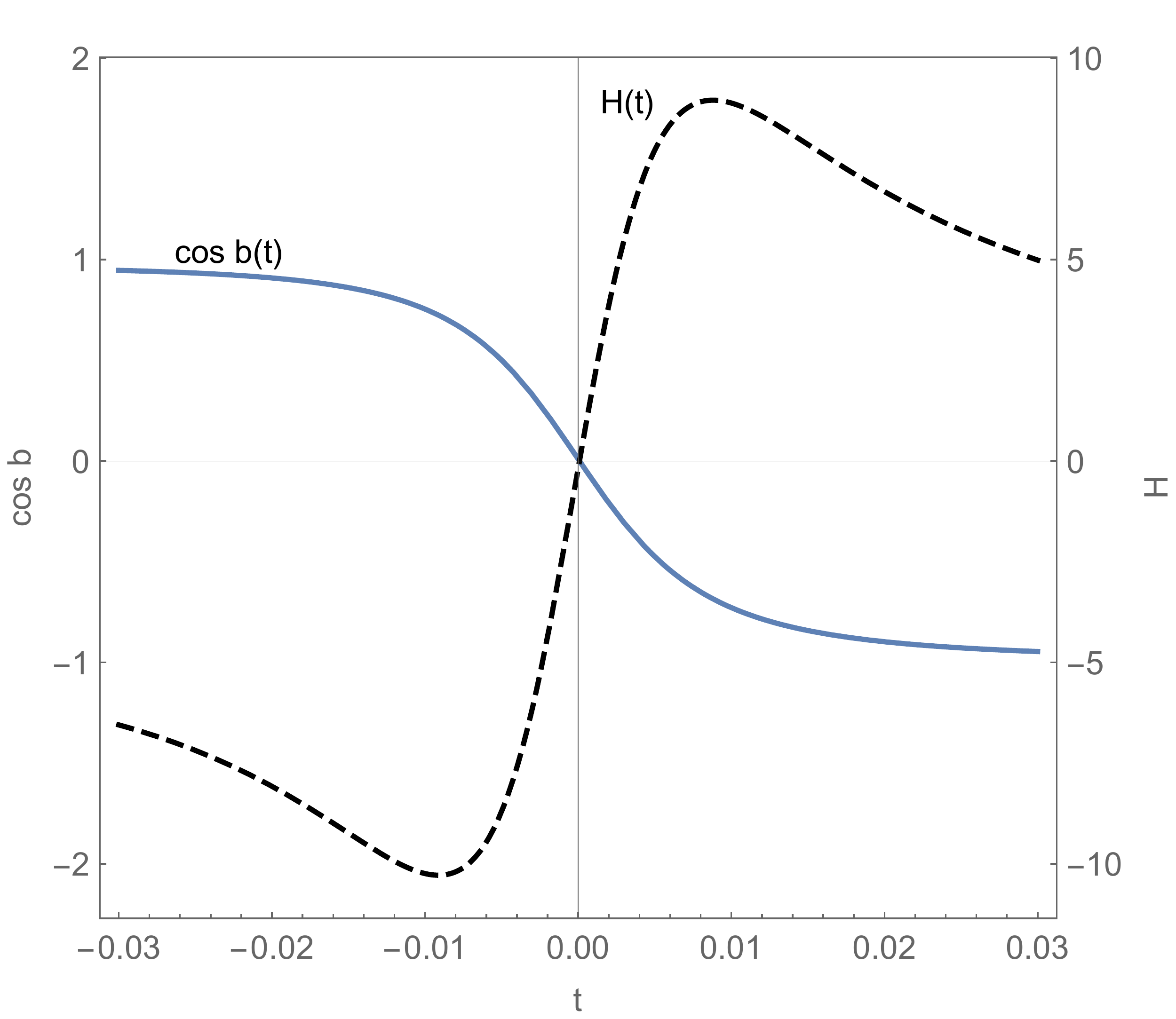}}
\captionsetup{justification=raggedright}
\caption{Fig. \ref{fig1a} shows the solutions of the equations of the model with classical initial conditions in the $\bar{\phi}-\dot{\bar{\phi}}$ diagram in the sector $\sin b<0$. For better clarity, we select $\lambda=\xi=1$. The dashed lines depict the evolution in the contraction phase of a universe. In the contraction phase, all solutions started from the origin in the remote past with classical initial conditions, and end up at the bounce surface shown by the external curve. As a representative,  the closed curve $a$ shows the complete solution for the initial condition $\phi=1\times10^{3}\kappa^{-\frac{1}{2}}$, $\dot{\phi}=1\times10^{2}\kappa$, in which the dashed line depicts the contraction phase and the solid line depicts the expansion phase. This solution originated from the remote past, passes through the bounce and evolves to the origin in the asymptotic future, and this solution also has an inflationary phase. Fig. \ref{fig1b} shows the evolution of $\cos b$ and $H$ around the bounce for the solution $a$. $t=0$ denotes the instant of bounce. The quantities are given by setting $\kappa=1$.}
\label{fig1}
\end{figure}

 To illustrate the behavior of the solutions, in Fig. \ref{fig1a} we show the trajectories of solutions of equations in the contraction phase of a universe in the sector $\sin b<0$ of the phase space.

Now, we derive the spectral indices of the slow-roll inflation in this context.  In the Jordan frame, the number of \emph{e}-folds from the moment at which $k=F^2(\bar{\phi})aH$ till the end of inflation is given by
 \ba
 &&N=\int^{t_e}_{t_i}Hdt
 =\int^{\bar{\phi}_i}_{\bar{\phi}_e}
 \frac{H}{\big(d\bar{\phi}/dt\big)}d\bar{\phi}
 \simeq\kappa\int^{\bar{\phi}_i}_{\bar{\phi}_e}
 \frac{G(\bar{\phi})V(\bar{\phi})}{4F(\bar{\phi})F'(\bar{\phi})V(\bar{\phi})+F^2(\bar{\phi})V'(\bar{\phi})}d\bar{\phi}\nn\\
 &&=\bigg[\frac{3}{16}\big(1+\xi\kappa\bar{\phi}^2\big)
 +\frac{1}{20\xi}\big(1+\xi\kappa\bar{\phi}^2\big)^{\frac{5}{2}}
 +\frac{1}{16}\ln\big(1+\xi\kappa\bar{\phi}^2\big)\bigg]\bigg|^{\bar{\phi}=\bar{\phi}_i}_{\bar{\phi}=\bar{\phi}_e},\label{NJor}
 \ea
 where $t_i$ denotes the moment at which $k=F^2(\bar{\phi})aH$, $t_e$ denotes the end of inflation, and $\bar{\phi}_i\equiv\bar{\phi}|_{t=t_i}$,  $\bar{\phi}_e\equiv\bar{\phi}|_{t=t_e}$.  The value of $\bar{\phi}_e$ can be derived from the condition $|\epsilon_1|=1$. Using the relation in Eq. (\ref{epsilons}), we can deduce that $\bar{\phi}_e$ is determined by
 \ba
 \frac{8+10\xi\kappa\bar{\phi}_e^2}{\kappa\bar{\phi}_e^2\big[\frac{3}{2}\xi^2\kappa\bar{\phi}_e^2 +(1+\xi\kappa\bar{\phi}_e^2)^{\frac{5}{2}}\big]}=1.
 \ea

Using Eqs. (\ref{PRs}), (\ref{nSs}), and (\ref{rs}), we derive the power spectrum and spectral index of the scalar perturbation and the tensor-to-scalar ratio,
\ba
&&P_{\mathcal{R}}\simeq\frac{\kappa^3\lambda}{768\pi^2}\frac{\kappa^3\bar{\phi}_i^6\big[\frac{3}{2}\xi^2\kappa\bar{\phi}_i^2 +(1+\xi\kappa\bar{\phi}_i^2)^{\frac{5}{2}}\big]}{(1+\xi\kappa\bar{\phi}_i^2)^2},\label{PRxi}\\
&&n_S-1\simeq-\frac{24\xi^3\kappa^2\bar{\phi}_i^4+48\xi^2\kappa\bar{\phi}_i^2+(24+28\xi\kappa\bar{\phi}_i^2)(1+\xi\kappa\bar{\phi}_i^2)^{\frac{5}{2}}}{\kappa\bar{\phi}_i^2\big[\frac{3}{2}\xi^2\kappa\bar{\phi}_i^2 +(1+\xi\kappa\bar{\phi}_i^2)^{\frac{5}{2}}\big]^2},\label{nSxi}\\
&&r\simeq \frac{128}{\kappa\bar{\phi}_i^2\big[\frac{3}{2}\xi^2\kappa\bar{\phi}_i^2 +(1+\xi\kappa\bar{\phi}_i^2)^{\frac{5}{2}}\big]}.\label{rxi}
\ea

From Eq. (\ref{NJor}), we find that in the case $\xi\gg (\frac{16N}{3})^\frac{3}{2}$, we have
\ba
1+\xi\kappa\bar{\phi}_i^2\simeq \frac{16}{3}N\bigg[1-\frac{4}{15}\sigma\bigg],\label{Nxiappro}
\ea
where $\sigma\equiv\frac{1}{\xi}\bigg(\frac{16N}{3}\bigg)^\frac{3}{2}$.
Substituting Eq. (\ref{Nxiappro}) into Eqs. (\ref{PRxi}), (\ref{nSxi}), and (\ref{rxi}), in the case $\xi\gg (\frac{16N}{3})^\frac{3}{2}$, we obtain
\ba
P_{\mathcal{R}}\simeq\frac{\lambda}{18\pi^2\xi^2}N^2
\bigg[1+\frac{7}{15}\sigma\bigg],\label{PRlxil}\quad
n_S-1\simeq-\frac{2}{N}\bigg[1+\frac{1}{10}\sigma\bigg],\quad
r\simeq\frac{3}{N^2}\bigg[1-\frac{2}{15}\sigma\bigg].
\ea

For $k=0.002$ Mpc$^{-1}$, the current observation gives
\ba
P_{\mathcal{R}}=(2.35\pm0.07)\times10^{-9},\quad n_S=0.9649\pm0.0042,\quad r<0.056,\label{obcons}
\ea
at $68\%$ C.L. \cite{Akrami:2018}.

\begin{figure}[h]
\centering
\includegraphics[height=5.5cm]{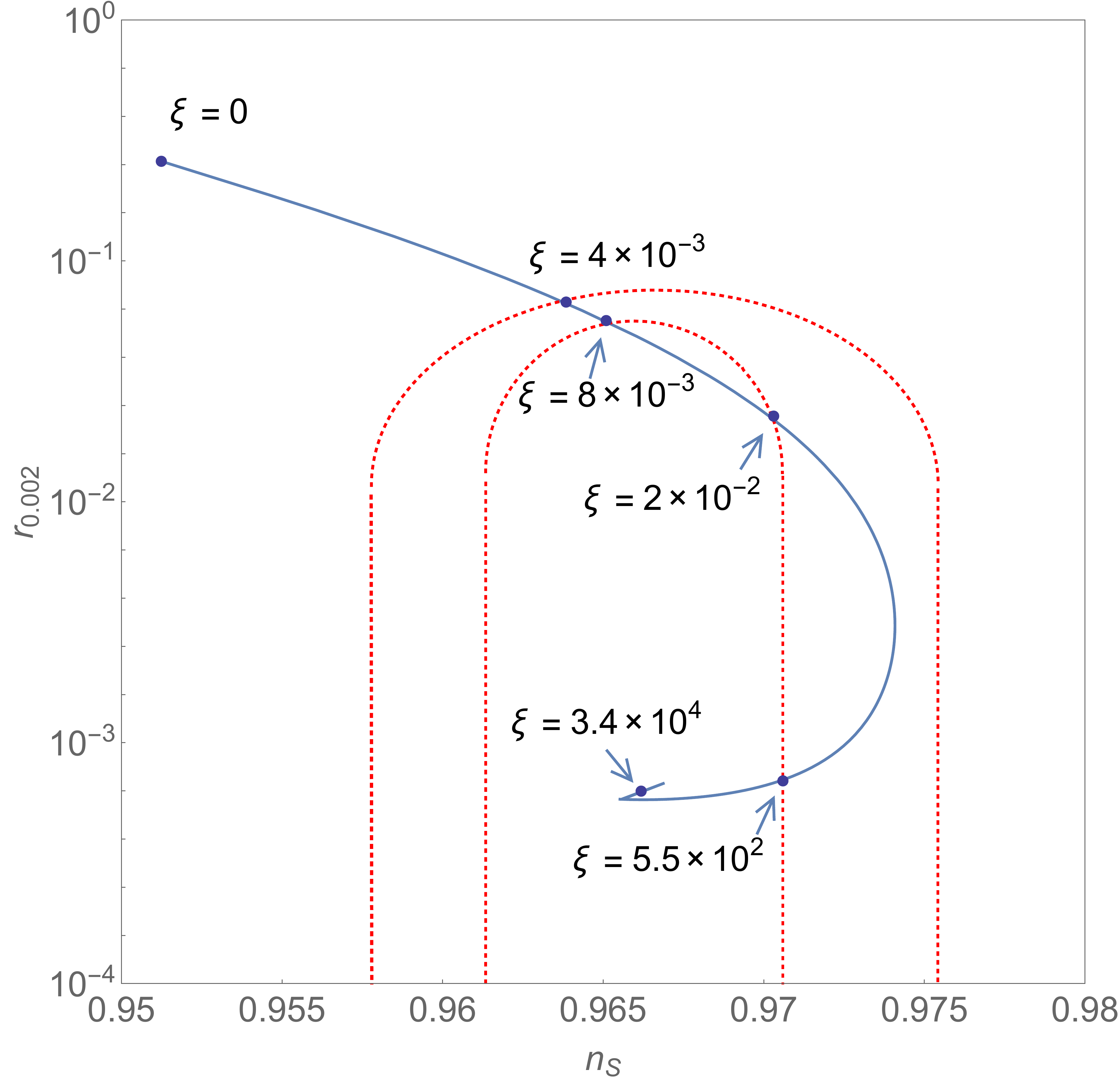}
\captionsetup{justification=raggedright}
\caption{Two dimensional observational constraints on the slow-roll inflation of the model in the ($n_S$, $r$) plane with the number of \emph{e}-folds $N=60$ and the wave number $k=0.002$ Mpc$^{-1}$ in the low-energy limit with $\cos b\rightarrow-1$. The dotted curves represent the $68\%$ C.L. (inside) and $95\%$ C.L. (outside) boundaries derived by the joint analysis of \textsl{Planck}2018+BK15+BAO. The solid curve show the theoretical predictions of the model  with the increase of $\xi$.}
\label{fig2}
\end{figure}

 Assuming that $N\simeq60$ for $k=0.002$ Mpc$^{-1}$,
in the limit $\xi\rightarrow\infty$, we have
\ba
\quad n_S\rightarrow 0.9667,\quad r\rightarrow 8.3\times10^{-4},
\ea
which is in complete agreement with the observation. In addition, for the Higgs field with the self-coupling parameter $\lambda\simeq0.13$, we obtain $\xi\simeq3.4\times10^{4}$, $n_S\simeq 0.966$, $r\simeq 8.1\times10^{-4}$. In Fig. \ref{fig2}, we show the theoretical predictions of $n_S$ and $r$ for different values of $\xi$.

The fact that the above result agrees well with the observation can be easily explained in the Einstein frame. In the large $\xi$ limit, we have
\ba
\bar{\tilde{\phi}}\equiv\int d\phi\frac{\sqrt{G\big(\bar{\phi}\big)}}{F\big(\phi\big)}
\simeq\frac{1}{\mu}\ln\big(1+\xi\kappa\bar{\phi}^2\big),
\ea
where $\mu\equiv\sqrt{\frac{8\kappa}{3}}$. The potential in the Einstein frame becomes
 \ba
 \tilde{V}\equiv F^4\big(\bar{\phi}\big)V\big(\bar{\phi}\big)\simeq \frac{1}{4\xi^2}\Big(1-e^{-\mu\bar{\tilde{\phi}}}\Big)^2,
 \ea
 which coincides with the $\alpha$-attractor potential favored by the current observation \cite{Kallosh:2013}.

 It is necessary to mention that two requirements should be satisfied in order to make the above analysis of the slow-roll inflation justified: 1. The slow-roll inflation satisfying the observational constraints in (\ref{obcons}) can take place in this model. 2. For $k\geq 0.002$ Mpc$^{-1}$, the energy density at $k=F^2(\bar{\phi})aH$ must be smaller than the critical energy density by many orders of magnitude in order that the Mukhanov equations (\ref{SMuk}) and (\ref{TMuk}) are applicable.

  Let us check whether the first requirement can be satisfied.  To find out which solutions can pass through the slow-roll inflation satisfying the observational constraints, considering that in the phase space every solution can reach the bounce surface, we can use  numerical analysis to check which phase space points on the bounce surface can lead to the desired slow-roll inflation. Note that there exists a symmetry of the phase space equations of motion: given a
solution \big($\bar{\phi}(t)$,$\dot{\bar{\phi}}(t)$\big) of the equations of motion, \big($-\bar{\phi}(t)$,$-\dot{\bar{\phi}}(t)$\big) is also a solution. Therefore, in the discussion of background dynamics it suffices to focus on $\dot{\bar{\phi}}_B>0$ and allow $\bar{\phi}_B$ to take both positive and negative values, where $(\bar{\phi}_B,\dot{\bar{\phi}}_B)$ represents the phase space points on the bounce surface.
Since the effective potential in the Jordan frame $V(\bar{\phi})=\frac{\lambda\bar{\phi}^4}{4(1+\xi\kappa\bar{\phi}^2)^{\frac{1}{2}}}$
 increases monotonically with respect to $\bar{\phi}$ for $\bar{\phi}>0$, for given $\xi$ and $\lambda$, $|\bar{\phi}_B|$  is bounded from above by $|\bar{\phi}_{max}|$ which is uniquely determined by
 \ba
 \frac{\lambda\bar{\phi}_{max}^4}{4(1+\xi\kappa\bar{\phi}_{max}^2)^{\frac{1}{2}}}=\rho_c.
 \ea
 We denote the value of $\bar{\phi}_B$ that can lead to the desired slow-roll inflation by $\bar{\phi}_B^{sl}$ and the set of $\bar{\phi}_B^{sl}$ by $\Omega$, then our task is to fix the range of $\Omega$.

 To check whether the second requirement can be satisfied, we can select the energy density $\rho_i\equiv\rho|_{k=F^2(\bar{\phi})aH}$ with $k=0.002$ Mpc$^{-1}$ and check whether the fraction $\frac{\rho_i}{\rho_c}$ is significantly smaller than $1$.

 In Table \ref{Table 1}, we list the numerical results for different $\xi$ and $
 \lambda$, from which we see that the two requirements can indeed be satisfied.
 \begin{center}
\begin{table}[h]
\begin{ruledtabular}
\begin{tabular}[c]{cccccc}
  $\xi$&$\lambda$&$\phi_{max}$&$\Omega$ & $\rho_i/\rho_c$ & $\cos b_i$ \\
  \hline
  $1\times10^{-2}$&$1.88\times10^{-12}$&$3.81\times10^{4}$ & $[-\phi_{max},-20.9]\cup[3.83,\phi_{max}]$ & $8.95\times10^{-9}$&$-1+4.48\times10^{-9}$ \\
  \hline
  $6\times10^{2}$&$4.76\times10^{-5}$&$8.11\times10^{2}$ & $[-\phi_{max},-2.02]\cup[-0.99,\phi_{max}]$ & $6.85\times10^{-8}$&$-1+3.43\times10^{-8}$ \\
  \hline
  $3.4\times10^{4}$&$0.13$&$1.14\times10^{2}$ & $[-\phi_{max},-0.81]\cup[-0.26,\phi_{max}]$ & $5.65\times10^{-10}$ &$-1+2.83\times10^{-10}$ \\
  \end{tabular}
\captionsetup{justification=raggedright}
\caption{Conditions for occurrence of the slow-roll inflation satisfying the constraints in (\ref{obcons}) and the value of $\rho_i/\rho_c$ for various $\xi$ and $
 \lambda$ in the case $\dot{\bar{\phi}}_B>0$. The value of the scalar field is given by setting $\kappa=1$ and $\cos b_i\equiv\cos b|_{k=F^2(\bar{\phi})aH}$ with $k=0.002$ Mpc$^{-1}$.}\label{Table 1}
\end{ruledtabular}
\end{table}
\end{center}

  We mention that the analysis in this section is based on the choice that a contracting universe is described by classical STT in the asymptotic past. If we ask that the background dynamics of a contracting universe is described by the low-energy limit of the $b_{-}$ branch of equations of motion in the asymptotic past, then after the bounce an expanding universe will be described by the $b_{+}$ branch of equations of motion, and numerical analysis shows that the slow-roll inflation satisfying the observational constraints cannot take place in this case.

\section{Summary and remarks}
The previous investigation of LQC of STT with the holonomy correction shows that in the cosmological case there exists two different branches of background equations of motion in the Jordan frame, i.e., the $b_{+}$ branch and the $b_{-}$ branch. In the low-energy limit, the  $b_{+}$ branch of equations reproduce the equations of classical STT  while the the $b_{-}$ branch of equations do not.  The evolution of an expanding universe can be described by either of the two branches. In this paper, we mainly study the cosmological dynamics of an expanding universe described by the $b_{-}$ branch of equations of motion and especially focus on the perturbation dynamics in the low-energy limit with $\cos b\rightarrow-1$ because it can provide important information of the holonomy correction even when the energy density is significantly lower than the Planck scale.  The main results obtained in this paper are summarized as follows.

First, using the method of dynamical analysis we show that the low energy limit with $\cos b\rightarrow-1$ can be a local attractor in the expansion phase of a universe, which means it is possible for the solutions of the background equations of motion to stably evolve to the low-energy limit with $\cos b\rightarrow-1$. Then, we derive the background Hamiltonian (\ref{BHFRW2}) which can yield the the background equations of motion in the limit $\cos b\rightarrow-1$, we also show that the background Hamiltonian (\ref{BHFRW2}) and the classical background Hamiltonian (\ref{BHFRW}) can be regarded as two different limiting cases of the background Hamiltonian (\ref{BHFRWLQC}) of LQC in the low-energy limit. Next, by imposing the anomaly-free condition we obtain a unique set of constraints which can yield a closed constraint algebra on the spatially flat FRW background in the low-energy limit. In particular, the constraint algebra (\ref{PoissonHam}) between two smeared Hamiltonian constraints explicitly shows that the spacetime structure is deformed by quantum corrections. In this way, we fix the anomaly-free Hamiltonian (\ref{Hamb-}). We also show that the constraints can be reexpressed in terms of the Ashtekar variables. Moreover, we find that using the field redefinitions in (\ref{canotrans}) the Hamiltonian (\ref{Hamb-}) can be rewritten in the form of the minimally coupled case, and this fact allows us to compare the physical results between the Jordan frame and the Einstein frame.

In the latter half of the paper, we mainly focus on the cosmological perturbations in this context and their applications in slow-roll inflation.  First, we expand the Hamiltonian (\ref{Hamb-}) to second order of perturbations and derive the canonical equations of the perturbed variables; then, we construct the gauge invariant perturbed variables and derive the second order evolution equations of the perturbed variables from the canonical equations. From these equations we learn that the propagation speed of the perturbations is subject to the quantum gravity effects. Nevertheless, it can be proved that the causality is still respected by quantum corrections.  Furthermore, we solve the Mukhanov equations under the slow-roll approximation and compare the results derived in the Jordan frame with those in the Einstein frame. It is found that to linear order of slow-roll parameters the power spectra and spectral indices in the Jordan frame coincide with those in the Einstein frame. Finally, we study a specific model of STT using the results obtained in the previous sections. We find that
in this model a contracting universe described by classical STT in the remote past can pass through the bounce and evolve into an expanding universe described by the $b_{-}$ branch of equations of motion, and finally it will approach the minimally coupled case in the asymptotic future. We also show that in this case the slow-roll inflation can take place and the spectral indices of the slow-roll inflation agree well with the latest astrophysical observations.

To summarize, in this paper, we have constructed an alternative consistent theory different from the classical STT in the low-energy limit of LQC in the Jordan frame, and the two theories can be regarded as different limiting cases of loop quantum STT.

At the end of this paper, we list some future research directions that the analysis in this paper can be extended to.

(1) The fact that the constraint algebra associated with the Hamiltonian constraint (\ref{Hamconstraint2}) is closed to arbitrary order of perturbations on the spatially flat FRW background allows us to perform the analysis involving higher order perturbations such as calculating the non-Gaussianity in the limit $\cos b\rightarrow-1$. Besides, it is also worth exploring whether the above result can be extended to other backgrounds with different topologies or symmetries.

(2) In the solution (\ref{svsk}) of Eq. (\ref{SMukk}), we choose $c_1(k)=1$, $c_2(k)=0$ for the comoving wave numbers $k\geq0.002$ Mpc$^{-1}$, which corresponds to selecting the Bunch-Davies vacuum in the small scale limit for these wave numbers. In this treatment, we are actually assuming that the preinflationary quantum gravity effects on these wave numbers can be neglected. To verify the justifiability of this assumption, in future work, we can derive the perturbation equations that are valid in the whole energy range of LQC and analyze the preinflationary quantum effects on these wave numbers.

(3) The fact that there exist two different effective theories in the low-energy limit of LQC has previously been pointed out in Ref. \cite{Assanioussi:2018} for the minimally coupled case; however, in Ref. \cite{Assanioussi:2018}, the authors used the modified holonomy quantization prescription, which is different from our case since we use the standard holonomy quantization prescription for the nonminimally coupled case in this paper. Recently, the modified holonomy quantization prescription has been extended to Brans-Dicke theory in Ref. \cite{Song:2020}, and it is interesting to generalize the results there to STT in the future research.

\begin{acknowledgements}
 The author thanks Dr. Long Chen for helpful discussions. This work is supported by NSFC (Grant No. 11905178) and Nanhu Scholars Program for Young Scholars of Xinyang Normal University.
\end{acknowledgements}
\appendix
\section{Derivation of the quantum correction term in Eq. (\ref{qcterm})}

The smeared background Hamiltonian in the low-energy limit with  $\cos b\rightarrow-1$ is expressed as
\ba
\textbf{H}^{(0)}_{b_{-}}=&&\int_{\Sigma}d^3x
\bar{N}\big[\mathcal{C}^{(0)}+\mathcal{Q}\big(a,p,\bar{\phi},\bar{\pi}\big)\big],
\ea
where the expression of $\mathcal{C}^{(0)}$ is in (\ref{C0}).

Using the Hamilton's equation and the commutation relation in (\ref{PBADMB}), we obtain the canonical equations of motion of the background variables,
\ba
\frac{da}{d\tau}&=&\bar{N}\Bigg[-\kappa p\frac{K\big(\bar{\phi}\big)}{G\big(\bar{\phi}\big)}
-\frac{\bar{\pi}}{2a^2}\frac{F'\big(\bar{\phi}\big)}{G\big(\bar{\phi}\big)}
+\frac{1}{6a}\frac{\partial \mathcal{Q}}{\partial p}\Bigg],\nn\\
\frac{dp}{d\tau}&=&\bar{N}\Bigg[-\frac{\kappa p^2}{2a} \frac{K\big(\bar{\phi}\big)}{G\big(\bar{\phi}\big)}-\frac{p\bar{\pi}}{a^3}
\frac{F'\big(\bar{\phi}\big)}{G\big(\bar{\phi}\big)}
+\frac{\bar{\pi}^2}{4a^5}\frac{F\big(\bar{\phi}\big)}{G\big(\bar{\phi}\big)}-\frac{1}{6a}\frac{\partial \mathcal{Q}}{\partial a}-\frac{a^2}{2}V(\bar{\phi})\Bigg],\nn\\
\frac{d\bar{\phi}}{d\tau}&=&\bar{N}\Bigg[-\frac{3p}{a}\frac{F'\big(\bar{\phi}\big)}{G\big(\bar{\phi}\big)}
+\frac{\bar{\pi}}{a^3}\frac{F\big(\bar{\phi}\big)}{G\big(\bar{\phi}\big)}+\frac{\partial \mathcal{Q}}{\partial \bar{\pi}}\Bigg],\nn\\
\frac{d\bar{\pi}}{d\tau}&=&\bar{N}\Bigg[3\kappa a p^2\Bigg(\frac{K\big(\bar{\phi}\big)}{G\big(\bar{\phi}\big)}\Bigg)'
+\frac{3p\bar{\pi}}{a}\Bigg(\frac{F'\big(\bar{\phi}\big)}{G\big(\bar{\phi}\big)}\Bigg)'
-\frac{\bar{\pi}^2}{2a^3}\Bigg(\frac{F\big(\bar{\phi}\big)}{G\big(\bar{\phi}\big)}\Bigg)'
-a^3V'\big(\bar{\phi}\big)-\frac{\partial \mathcal{Q}}{\partial \bar{\phi}}\Bigg].\label{Hameq}
\ea
From these equations, we find that on the constraint surface $\mathcal{C}^{(0)}_{b_{-}}+\mathcal{Q}=0$ the Friedmann equation and Klein-Gordon equation read, respectively, as
\ba
&&\bigg(F\big(\bar{\phi}\big)H-\frac{1}{2}\dot{F}\big(\bar{\phi}\big)\bigg)^2
=\frac{\kappa}{3}\rho_e+\frac{\kappa}{3}\mathcal{X}\big(a,p,\bar{\phi},\bar{\pi}\big)-\frac{\kappa p}{a}\mathcal{Y}\big(a,p,\bar{\phi},\bar{\pi}\big)+\mathcal{Y}^2\big(a,p,\bar{\phi},\bar{\pi}\big),\label{Appeom1}\\
&&\ddot{\bar{\phi}}+3H\dot{\bar{\phi}}+\frac{1}{2}\frac{\dot{G}\big(\bar{\phi}\big)}{G\big(\bar{\phi}\big)}
\dot{\bar{\phi}} -\frac{2F'\big(\bar{\phi}\big)V\big(\bar{\phi}\big)
-F\big(\bar{\phi}\big)V'\big(\bar{\phi}\big)}{G\big(\bar{\phi}\big)}
+\frac{1}{a^6F\big(\bar{\phi}\big)\sqrt{G\big(\bar{\phi}\big)}}\mathcal{Z}\big(a,p,\bar{\phi},\bar{\pi}\big)=0,\label{Appeom2}
\ea
where
\ba
\mathcal{X}\big(a,p,\bar{\phi},\bar{\pi}\big)
&=&\frac{F'\big(\bar{\phi}\big)3a^2p-F\big(\bar{\phi}\big)\bar{\pi}}{a^3}\frac{\partial \mathcal{Q}}{\partial \bar{\pi}}
-\frac{1}{2}G\big(\bar{\phi}\big)\bigg(\frac{\partial \mathcal{Q}}{\partial \bar{\pi}}\bigg)^2
+\frac{F\big(\bar{\phi}\big)\mathcal{Q}}{a^3},\nn\\
\mathcal{Y}\big(a,p,\bar{\phi},\bar{\pi}\big)
&=&\frac{\Big(F'\big(\bar{\phi}\big)\Big)^23a^2p
-F\big(\bar{\phi}\big)F'\big(\bar{\phi}\big)\bar{\pi}}{a^3G\big(\bar{\phi}\big)}
-\frac{F'\big(\bar{\phi}\big)}{2}\frac{\partial \mathcal{Q}}{\partial \bar{\pi}}
+\frac{F\big(\bar{\phi}\big)}{6a^2}\frac{\partial \mathcal{Q}}{\partial p},\nn\\
\mathcal{Z}\big(a,p,\bar{\phi},\bar{\pi}\big)
&=&\Bigg\{\frac{F'\big(\bar{\phi}\big)3a^2p-F\big(\bar{\phi}\big)\bar{\pi}}{\sqrt{G\big(\bar{\phi}\big)}},
a^3F\big(\bar{\phi}\big)\mathcal{Q}\Bigg\}
-\Bigg\{a^3\sqrt{G\big(\bar{\phi}\big)}\frac{\partial \mathcal{Q}}{\partial \bar{\pi}},a^3F\big(\bar{\phi}\big)\mathcal{C}^{(0)}\Bigg\}.
\ea
Comparing Eqs. (\ref{Appeom1}) and (\ref{Appeom2}) with Eqs. (\ref{eomFbminus}) and (\ref{eomKbminus}), we find the following equations must be satisfied:
\ba
&&\frac{\kappa}{3}\mathcal{X}\big(a,p,\bar{\phi},\bar{\pi}\big)-\frac{6\kappa p}{6a}\mathcal{Y}\big(a,p,\bar{\phi},\bar{\pi}\big)+\mathcal{Y}^2\big(a,p,\bar{\phi},\bar{\pi}\big)=0,\nn\\
&&\frac{1}{a^6F\big(\bar{\phi}\big)\sqrt{G\big(\bar{\phi}\big)}}\mathcal{Z}\big(a,p,\bar{\phi},\bar{\pi}\big)
=\frac{6F'\big(\bar{\phi}\big)V\big(\bar{\phi}\big)}{G\big(\bar{\phi}\big)},
\ea
which together with the constraint $\mathcal{C}^{(0)}_{b_{-}}+\mathcal{Q}=0$ give the following solution:
\ba
\mathcal{Q}\big(a,p,\bar{\phi},\bar{\pi}\big)=(1+A)\frac{6pF'\big(\bar{\phi}\big)\bar{\pi}}{aG\big(\bar{\phi}\big)}
+A\mathcal{C}^{(0)},
\ea
where $A$ is an arbitrary constant.
In the case $F(\bar{\phi})=K(\bar{\phi})=1$, the background Hamiltonian should reduce to the background Hamiltonian of the minimally coupled case, which gives $A=0$. Thus, we obtain
$\mathcal{Q}\big(a,p,\bar{\phi},\bar{\pi}\big)=\frac{6pF'\big(\bar{\phi}\big)\bar{\pi}}{aG\big(\bar{\phi}\big)}$.

\end{document}